\documentclass[a4paper,12pt,nofootinbib]{article}
\usepackage{latexsym}
\usepackage{epsfig}
\usepackage[usenames]{color}
\usepackage{amsmath}
\usepackage{amsfonts}
\usepackage{amssymb}

\def\){\right)}
\def\({\left( }
\def\]{\right] }
\def\[{\left[ }

\def\no{\nonumber \\}

\def\be{\begin{equation}}
\def\ee{\end{equation}}
\def\ba{\begin{eqnarray}}
\def\ea{\end{eqnarray}}
\def\no{\nonumber \\}

\def\ra{\rangle}
\def\la{\langle}

\newcommand {\cM}{{\cal M}}
\newcommand {\cN}{{\cal N}}

\def\la{\langle}
\def\ra{\rangle}

\begin{document}
\begin{titlepage}
\begin{center}
 {\Large \bf Walls of massive K\"ahler sigma models\\
\vskip 0.2cm
  on $SO(2N)/U(N)$ and $Sp(N)/U(N)$}

\vskip 0.8cm
\normalsize
\renewcommand\thefootnote{\alph{footnote}}

\vskip 0.5cm

{\bf Masato Arai$^\dagger$\footnote{Masato.Arai(at)utef.cvut.cz}
 and Sunyoung Shin$^\sharp$\footnote{shin(at)theor.jinr.ru}
}

\vskip 0.5cm

{\it $^\dagger$ Institute of Experimental and Applied Physics,
   Czech Technical University in Prague,
   Horsk\' a 3a/22, 128 00 Prague 2, Czech Republic\\
   \vskip 0.5cm
 $^\sharp$ Bogoliubov  Laboratory of Theoretical Physics, JINR,}\\
{\it 141980, Dubna, Moscow region, Russia} \\

\vskip 2cm

\begin{abstract}
We study the Bogomol'nyi-Prasad-Sommerfield wall solutions in
massive K\"ahler nonlinear sigma models on $SO(2N)/U(N)$ and
$Sp(N)/U(N)$ in three-dimensional spacetime. We show that
$SO(2N)/U(N)$ and $Sp(N)/U(N)$ models have $2^{N-1}$ and $2^N$
discrete vacua, respectively. We explicitly construct the exact BPS
multiwall solutions for $N\le 3$.
\end{abstract}
\end{center}
\end{titlepage}

\renewcommand{\theequation}{\thesection.\arabic{equation}}
\newpage
\section{Introduction}
\setcounter{equation}{0}
Topological solitons in supersymmetric (SUSY) theories preserve a
fraction of the original SUSY \cite{Witten:1978mh} as they saturate
the Bogomol'nyi-Prasad-Sommerfield (BPS) bound on the energy
\cite{BPS}. Domain walls are one of the simplest BPS objects, which
conserve half of SUSY \cite{de Azcarraga:1989gm, CQR,
Abraham:1992vb}.

A systematic method to construct domain wall solutions in SUSY
$U(N_C)$ gauge theories coupled to $N_F~(N_F>N_C)$ massive
hypermultiplets in the fundamental representation in the presence of
the Fayet-Iliopoulos has been proposed
\cite{Isozumi:2004jc,Isozumi:2004va}. The mass term forms nontrivial
scalar potential, yielding ${}_{N_F}C_{N_C}$ discrete vacua. Exact
domain wall solutions interpolating these vacua have been obtained.
This construction method is called the moduli matrix approach.
Subsequently this approach has been applied to obtain other types of
solitonic solutions such as monopole-vortex-wall systems
\cite{INOS3}, domain wall webs \cite{EINOS1}, non-Abelian vortices
\cite{EINOS2}, instanton-vortex systems \cite{EINOS3} and Skyrmions
\cite{ENOT}. For a comprehensive review, see \cite{INOS2}. The model
considered in \cite{Isozumi:2004jc,Isozumi:2004va} has an
interesting limit: Taking infinite gauge coupling limit, the kinetic
terms of gauge fields and their superpartners are vanishing and they
turn out to be just a Lagrange multiplier, yielding constraints to
hypermultiplets. The action with such a limit gives the quotient
action of massive hyper-K\"ahler (HK) nonlinear sigma model (NLSM)
on cotangent bundle over Grassmannian, $T^*G_{N_F,N_C}$. Vacuum
structure of this NLSM was originally addressed in \cite{ANS},
giving the same number of vacua before taking infinite gauge
coupling limit while domain wall solution was obtained only in
simple case, for instance, $T^*G_{2,1}\simeq T^*{\bf C}P^1$
\cite{Abraham:1992vb,Gauntlett:2000bd,Gauntlett:2000de,Arai:2002xa,Arai:2003my}
until the moduli matrix approach was proposed.

The Grassmann manifold is one of the compact Hermitian symmetric
spaces (HSS), which we denote $\cM$. It consists of the four
classical types, the Grassmann manifold $G_{N+M,M}$, complex quadric
surface $Q^N=SO(N+2)/[SO(N)\times U(1)]$, $SO(2N)/U(N)$,
$Sp(N)/U(N)$ and two exceptional types, $E_6/[SO(10)\times U(1)]$
and $E_7/[E_6\times U(1)]$. It is interesting to investigate vacuum
structure and domain wall solutions in HK NLSM on cotangent bundle
over $\cM$ other than $G_{N+M,M}$ since the other HSS are expected
to give rich vacuum structure as well as abundant wall solutions
similar to HK NLSM on $T^*G_{N+M,M}$. The actions of HK NLSM on
cotangent bundle over classical HSS \cite{KG1,KG2,AKL1} \footnote{A
massless HK NLSM on the tangent bundle over the complex quadric
surface being one of the classical HSSs has been worked out in
\cite{AN}.} and on $T^*E_6/[SO(10)\times U(1)]$ \cite{AKL2} have
been obtained in projective superspace \cite{KLR,LR}, but they are
not the quotient actions. They are written in terms of physical
degrees of freedom without Lagrange multiplier and therefore they
are not gauge theories with infinite gauge coupling limit. The
moduli matrix approach can be easily applied to a quotient action,
but it is difficult to construct a quotient action on cotangent
bundle over $\cM$ except Grassmann manifold.

In \cite{Isozumi:2004jc,Isozumi:2004va}, it has been also shown that
when considering wall solutions in massive HK NLSM on $T^*
G_{N_F,N_C}$ the cotangent part is trivial. Only field coordinates
parameterizing base manifold have nontrivial configuration. It means
that we can simply drop the cotangent bundle part in massive HK
NLSM, reducing the massive HK NLSM to a simpler model, massive
K\"ahler NLSM.

In this paper, we investigate discrete vacua and domain wall
solutions interpolating them in massive K\"ahler NLSM on the HSS,
$SO(2N)/U(N)$ and $Sp(N)/U(N)$. These massive NLSMs are obtained
from the massless K\"ahler NLSM on $SO(2N)/U(N)$ and $Sp(N)/U(N)$ in
four-dimensional spacetime which are described as a quotient action
\cite{HN} by dimensional reduction. The actions possess a nontrivial
scalar potential which includes mass terms characterized by the
common Cartan matrices of $SO(2)$ and $Sp(N)$, leading to many
discrete vacua. We show that the $SO(2N)/U(N)$ model possesses
$2^{N-1}$ number of discrete vacua while $2^N$ number of discrete
vacua exists in the $Sp(N)/U(N)$ model. We derive wall solutions
interpolating those vacua and also discuss their properties.

Organization of this paper is as follows. In Section 2, we present
the massive NLSMs on $SO(2N)/U(N)$ and $Sp(N)/U(N)$. We also discuss
the discrete vacua induced by the mass term. In Section 3, we derive
half BPS equations. In Section 4 and 5, the exact wall solutions are
obtained in two models based on the moduli matrix approach. Section
6 is devoted to conclusion and discussion. In Appendix A, we explain
the vacuum structure of massive NLSMs on $SO(6)/U(3)$ and
$Sp(3)/U(3)$. In Appendix B, we show domain wall solutions in a
massive K\"ahler NLSM on ${\bf C}P^3$.
%
%
\section{Massive K\"ahler NLSM on $SO(2N)/U(N)$ and $Sp(N)/U(N)$} \label{sec;massiveKahler}
\setcounter{equation}{0}
In this section, we construct massive K\"ahler NLSMs on
$SO(2N)/U(N)$ and $Sp(N)/U(N)$ in three-dimensional spacetime by
dimensional reduction from the massless model in four-dimensional
spacetime. We follow the notation of \cite{WB}.

SUSY gauge theory inducing a quotient action of massless NLSMs on
$SO(2N)/U(N)$ and $Sp(N)/U(N)$ are formulated in terms of $\cN=1$
superfields \cite{HN}. These actions are obtained by imposing
quadric constraints on K\"ahler NLSM on the Grassmann manifold
$G_{2N,N}=\frac{SU(2N)}{SU(N)\times U(N)}$. $G_{2N,N}$ can be
constructed by an $N\times 2N$ matrix chiral superfield
$\phi_a^{~i}(x,\theta,\bar \theta) \; (a=1,\cdots,N,~i=1,\cdots,2N)$
and an $N\times N$ matrix vector superfield $V_a^{~b} (x,\theta,\bar
\theta)$ in the adjoint representation of $U(N)$. We introduce an
$N\times N$ matrix chiral superfield $\phi_0^{ab}(x,\theta,\bar
\theta)$ as an auxiliary field to impose the $F$-term constraint.
The Lagrangian of massless NLSMs on $SO(2N)/U(N)$ and $Sp(N)/U(N)$
is
\begin{eqnarray}
 {\cal L} = \int d^4 \theta
 (\phi_a^{~i}\bar{\phi}_i^{~b} (e^V)_b^{~a} - r^2 V_a^{~a})
 + \left(\int d^2 \theta \, \phi_0^{ab} (\phi_b^{~i} J_{ij}\phi^{{\mathrm{T}}j}_{~~a}) + {\rm
 c.c.}\right),\label{eq:lag_susy_4d}
\end{eqnarray}
where $r^2$ is the Fayet-Iliopoulos parameter and the matrix $J$ is
defined by
\begin{eqnarray}
 J=\mathbf{1}\otimes\left(
  \begin{array}{cc}
   0 & 1  \\
   \epsilon & 0 \\
  \end{array}
 \right)\,,
 ~~~~~ \epsilon=\left\{
 \begin{array}{cc}
 +1&~~SO(2N)/U(N)\\
 -1&~~Sp(N)/U(N).
 \end{array}
 \right.
 \label{eq:sosp_constraint}
\end{eqnarray}
All the repeated indices are implicitly summed over. The auxiliary
chiral superfield $\phi_0^{ab}(x,\theta,\bar \theta)$ is in a
symmetric (anti-symmetric) rank 2 tensor representation of $SU(N)$
\begin{eqnarray}
\phi_0^T=\epsilon\phi_0,
\end{eqnarray}
with $U(1)(\in U(N))$ charge $-2$. The $D$-term in
(\ref{eq:lag_susy_4d}) gives a quotient action of K\"ahler NLSM on
$G_{2N,N}$ while the $F$-term gives constraints, realizing
submanifolds $SO(2N)/U(N)$ and $Sp(N)/U(N)$ as quadrics of
$G_{2N,N}$. The latter is obtained by the equation of motion of
$\phi_0$ as
\begin{eqnarray}
 \phi_b^{~i} J_{ij}\phi^{{\mathrm{T}}j}_{~~a}=0\,.
\label{const}
\end{eqnarray}
For $\epsilon=+1$ case, the corresponding submanifold satisfying
(\ref{const}) has not only $SO(2N)$ symmetry but also parity in
flavor indices as $J$ in (\ref{eq:sosp_constraint}) is invariant
under $O(2N)$ group. The parity will be removed. We will discuss
this in detail in later sections.

Next we derive massive K\"ahler NLSMs on $SO(2N)/U(N)$ and
$Sp(N)/U(N)$ from (\ref{eq:lag_susy_4d}). As solitonic objects are
the interest, we consider only the bosonic part of
(\ref{eq:lag_susy_4d}) in the following context. We replace the
superfields of the Lagrangian with fields expanded by bosonic
components
\begin{eqnarray}
\phi_a^{~i}(x,\theta,\bar{\theta})&=&\phi_a^{~i}(x)+\theta^2F_a^{~i},\nonumber\\
\phi_0^{ab}(x,\theta,\bar{\theta})&=&\phi_0^{ab}(x)+\theta^2F_0^{ab},\nonumber\\
V_a^{~b}(x,\theta,\bar{\theta})&=&2\theta\sigma^\mu\bar{\theta}v_\mu+\frac{1}{2}\theta^2\bar{\theta}^2D,
\end{eqnarray}
and then we obtain
\begin{eqnarray}
{\mathcal{L}}_{\mathrm{bos}~4D}&=&-|D_\mu\phi_a^{~i}|^2+|F_a^{~i}|^2+\frac{1}{2}(D_a^{~b}\phi_b^{~i}\bar{\phi}_i^{~a}-D_a^{~a})     \nonumber\\
&&+\Big((F_0)^{ab}\phi_b^{~i}J_{ij}\phi^{\mathrm{T}j}_{~~a}+(\phi_0)^{ab}F_b^{~i}J_{ij}\phi^{\mathrm{T}j}_{~~a}
+(\phi_0)^{ab}\phi_b^{~i}J_{ij}F^{\mathrm{T}j}_{~~a}
+{\mathrm c.\mathrm c.}\Big),\label{eq:lag_4d}
\end{eqnarray}
where the Greek letter $\mu$ denotes a four-dimensional spacetime
index. We introduce the mass term by dimensional reduction along the
$x^3$-direction as follows
\begin{eqnarray}
\frac{\partial\phi_a^{~i}}{\partial x^3}=i\phi_a^{~j}M_j^{~i},~~~~~
\frac{\partial\bar{\phi}_i^{~a}}{\partial
x^3}=-i\bar{M}_i^{~j}\bar{\phi}_j^{~a}, \label{eq:killing_vec}
\end{eqnarray}
where
\begin{eqnarray}
M_j^{~i}=\mathrm{diag}(m_1,m_2,\cdots,m_N)\otimes\sigma_3.
\label{eq:mass}
\end{eqnarray}
$M_i^{~j}$ is the Cartan matrix of $SO(2N)$ and $Sp(N)$.
The components $m_i~(i=1,\cdots,N)$ are real and positive parameters
with a condition $m_i>m_{i+1}$. The mass term breaks the global
symmetries to $SO(2)^N$ and $Sp(1)^N$ for each model. We substitute
(\ref{eq:killing_vec}) into (\ref{eq:lag_4d}) to obtain the
Lagrangian for massive $SO(2N)/U(N)$ and $Sp(N)/U(N)$ in three
dimensions
\begin{eqnarray}
{\mathcal{L}}_{\mathrm{bos}~3D}&=&-|D_m\phi_a^{~i}|^2-|i\phi_a^{~j}M_j^{~i}-i\Sigma_a^{~b}\phi_b^{~i}|^2
+|F_a^{~i}|^2+\frac{1}{2}(D_a^{~b}\phi_b^{~i}\bar{\phi}_i^{~a}-D_a^{~a})     \nonumber\\
&&+\Big((F_0)^{ab}\phi_b^{~i}J_{ij}\phi^{\mathrm{T}j}_{~~a}+(\phi_0)^{ab}F_b^{~i}J_{ij}\phi^{\mathrm{T}j}_{~~a}
+(\phi_0)^{ab}\phi_b^{~i}J_{ij}F^{\mathrm{T}j}_{~~a}
+{\mathrm c.\mathrm c.}\Big), \label{3d-lag}
\end{eqnarray}
where $\Sigma=v_3$. A Roman letter index $m$ refers to the first
three components of the four-dimensional index $\mu$. The
constraints of the Lagrangian are
\begin{eqnarray}
&&\phi_a^{~i}\bar{\phi}_i^{~b}-\delta_a^{~b}=0,\label{eq:constraint1}\\
&&\phi_a^{~i}J_{ij}\phi^{\mathrm{T}j}_{~~b}=0.\label{eq:constraint2}
\end{eqnarray}
Eliminating the auxiliary fields $F_a^{~i}$ by its equation of
motion
\begin{eqnarray}
F_a^{~i}=-2(\bar{\phi}_0)_{ab}\phi^{\ast b}_{~~j}
{J}^{ji},
\end{eqnarray}
we obtain the following scalar potential
\begin{eqnarray}
V&=&|i\phi_a^{~j}M_j^{~i}-i\Sigma_a^{~b}\phi_b^{~i}|^2+4|(\phi_0)^{ab}\phi_b^{~i}|^2.
\end{eqnarray}
The vacuum condition is readily read off as
\begin{eqnarray}
 &&\phi_a^{~j}M_j^{~i}-i\Sigma_a^{~b}\phi_b^{~i}=0, \label{vac1} \\
 &&(\phi_0)^{ab}\phi_b^{~i}=0, \label{vac2}
\end{eqnarray}
with the constraints (\ref{eq:constraint1}) and
(\ref{eq:constraint2}). The condition (\ref{vac2}) gives
$(\phi_0)^{ab}=0$ or $\phi_a^{~i}=0$, but the latter solution is
inconsistent with (\ref{eq:constraint1}). Taking account of the
former, the vacuum condition is reduced to
\begin{eqnarray}
&&(m_k-\Sigma_a)\phi_a^{~2k-1}=0, \label{eq:vac_equation1} \\
&&(m_k+\Sigma_a)\phi_a^{~2k}=0, \label{eq:vac_equation2}
\end{eqnarray}
where we have used (\ref{eq:mass}). We have diagonalized $\Sigma$ by
using $U(N)$ gauge symmetry as
\begin{eqnarray}
 \Sigma={\rm diag}(\Sigma_1,\Sigma_2,\cdots,\Sigma_N).
 \label{eq:sigma}
\end{eqnarray}
The indices $k,a=\{1,\cdots,N\}$ are not summed over.

We solve the equations (\ref{eq:vac_equation1}) and
(\ref{eq:vac_equation2}) with the constraints (\ref{eq:constraint1})
and (\ref{eq:constraint2}). Above equations yield $\Sigma_a=\pm m_k$
for some combinations of $a$ and $k$. First let us consider a
solution $\Sigma_{a_1}=m_{k_1}$ with $a_1=k_1$ for some $a_1(k_1)$.
It leads to $\phi_{a_1}^{~2k_1-1}\neq 0$ and also
$\phi_{a_1}^{~2k_1}=0$ from (\ref{eq:vac_equation2}). Similarly,
considering a solution $\Sigma_{a_1}=-m_{k_1}$ with $a_1=k_1$ for
some $a_1(k_1)$, we have $\phi_{a_1}^{~2k_1-1}=0$ and
$\phi_{a_1}^{~2k_1}\neq 0$. For another gauge index $a_2(\neq a_1)$,
possible solutions are $\Sigma_{a_2}=\pm m_{k_2}$ with $k_1=k_2$ and
$\Sigma_{a_2}=\pm m_{k_2}$ with $k_1 \neq k_2$. However, the former
is not a solution since it forms vacuum expectation value $\phi$ not
satisfying the constraint (\ref{eq:constraint1}) or
(\ref{eq:constraint2}) while the latter forms $\phi$ satisfying the
constraints. Similarly, for $a_3$ we find that $\Sigma_{a_3}=\pm
m_{k_3}$ with $k_1 \neq k_3$ and $k_1 \neq k_2$ forms $\phi$
satisfying the constraints. Repeating the same discussion for the
other gauge indices, we have the following $\Sigma$ for the vacua
\begin{eqnarray}
(\Sigma_{a_1}, \Sigma_{a_2}, \Sigma_{a_3},\cdots,\Sigma_{a_N})
=(\pm m_{k_1}, \pm m_{k_2}, \pm m_{k_3}, \cdots, \pm m_{k_N}), \label{vac_sigma}
\end{eqnarray}
where $a_1=k_1$ with $a_i\neq a_j$ and $k_i \neq k_j$ for $i\neq j$.
The $U(N)$ gauge symmetry allows us to take (\ref{vac_sigma}) to the following form
\begin{eqnarray}
(\Sigma_{1}, \Sigma_{2}, \Sigma_{3},\cdots,\Sigma_{N}) =(\pm m_{1},
\pm m_{2}, \pm m_{3}, \cdots, \pm m_{N}). \label{vac_sigma2}
\end{eqnarray}
From this expression, we see that the number of possible solutions
is $2^{N}$. Since $\epsilon=+1$ in (\ref{eq:sosp_constraint})
defines $O(2N)$ group the half of the solutions are related by
parity to the other half. The number of vacua in $SO(2N)/U(N)$ is
therefore $2^{N-1}$. The number of vacua in $Sp(N)/U(N)$ is $2^N$ as
shown in (\ref{vac_sigma2}).

We consider $N=2$ case explicitly. There are four solutions
satisfying (\ref{eq:vac_equation1}) and (\ref{eq:vac_equation2})
with (\ref{eq:constraint1}) and (\ref{eq:constraint2}):
\begin{eqnarray}
&& \Phi_{\la 1 \ra}=\left(
  \begin{array}{cccc}
   1 & 0 & 0 & 0 \\
   0 & 0 & \alpha_1 & 0\\
  \end{array}
 \right),~~~~(\Sigma_1,\Sigma_2)=(m_1,m_2),\nonumber \\
&&\Phi_{\la 2 \ra}=\left(
  \begin{array}{cccc}
   1 & 0 & 0 & 0 \\
   0 & 0 & 0 & \alpha_2\\
  \end{array}
 \right),~~~~(\Sigma_1,\Sigma_2)=(m_1,-m_2),\nonumber\\
&& \Phi_{\la 3 \ra}=\left(
  \begin{array}{cccc}
   0 & 1 & 0 & 0 \\
   0 & 0 & \alpha_3 & 0\\
  \end{array}
 \right),~~~~(\Sigma_1,\Sigma_2)=(-m_1,m_2),\nonumber\\
&& \Phi_{\la 4 \ra}=\left(
  \begin{array}{cccc}
   0 & 1 & 0 & 0 \\
   0 & 0 & 0 & \alpha_4\\
  \end{array}
 \right),~~~~(\Sigma_1,\Sigma_2)=(-m_1,-m_2), \label{vacO4}
\end{eqnarray}
where $\la 1 \ra, \cdots, \la 4 \ra$ denote the labels of the vacua.
$\alpha_i=1~(i=1,\cdots,4)$ for $SO(4)/U(2)$ whereas $\alpha_i=\pm
1$ for $Sp(2)/U(2)$.

For $\epsilon=+1$, the parity in (\ref{vacO4}) can be identified by
$O(4)$ group elements. We define a rotation transformation
${\mathcal{R}}$ and a parity transformation ${\mathcal{P}}$ of
$O(4)$ group as
\begin{eqnarray}
{\mathcal{R}}=\left(
\begin{array}{cc}
 P & 0 \\
 0 & P \\
\end{array}
\right),~~~ {\mathcal{P}}=\left(
\begin{array}{cc}
 I & 0 \\
 0 & P \\
\end{array}
\right), \label{eq:PhiO4}
\end{eqnarray}
where {\footnotesize{$P=\left(
\begin{array}{cc}
 0 & 1 \\
 1 & 0 \\
\end{array}
\right)$}} and $I$ is a two-by-two identity matrix.
The vacua are then related by
\begin{eqnarray}
\Phi_{\la 1 \ra}=\Phi_{\la 4 \ra}{\mathcal{R}},~~
\Phi_{\la 2 \ra}=\Phi_{\la 3 \ra}{\mathcal{R}},~~
\Phi_{\la 1 \ra}=\Phi_{\la 2 \ra}{\mathcal{P}}.
\end{eqnarray}
It shows that (\ref{3d-lag}) with $\epsilon=+1$ involves two massive
$SO(4)/U(2)$ models related by parity. Two sets of two vacua $(\la 1
\ra, \la 4 \ra)$ and $(\la 2 \ra, \la 3 \ra)$ belong to each massive
$SO(4)/U(2)$ model. Therefore there exist two discrete vacua in a
single massive $SO(4)/U(2)$ model. $SO(4)/U(2)$ is isomorphic to
$\mathbf{C}P^1$ \cite{Higashijima:2001vk}. A massive K\"ahler NLSM
on $\mathbf{C}P^1$ has only two discrete vacua \cite{Arai:2003tc}.
The results are consistent.

For $\epsilon=-1$, (\ref{eq:constraint2}) defines an invariant
submanifold under the action of $Sp(2)$, leading to a single massive
$Sp(2)/U(2)$ model. We find that there are $2^2=4$ discrete vacua in
this case.
%
%
\section{BPS equations} \label{sec:bps}
\setcounter{equation}{0}
In this section, we derive the BPS equation for wall solutions from
the Bogomol'nyi completion of the Hamiltonian. We assume that fields
are static and all the fields depend only on the $x_1\equiv x$
coordinate. We also assume Poincar\'{e} invariance on the
two-dimensional world volume of walls to set $v_0=v_2=0$. The energy
along the $x$-direction is
\begin{eqnarray}
E&=&\int dx \Big(
|D\phi_a^{~i}|^2+|\phi_a^{~j}M_j^{~i}-\Sigma_a^{~b}\phi_b^{~i}|^2
+4|(\phi_0)^{ab}\phi_b^{~i}|^2\Big)\nonumber\\
&=&\int dx
\Big(|D\phi_a^{~i}\mp(\phi_a^{i}M_j^{~i}-\Sigma_a^{~b}\phi_b^{~i})|^2
+4|(\phi_0)^{ab}\phi_b^{~i}|^2\Big)\pm T \nonumber \\
&\geq & \pm T,
\end{eqnarray}
with the constraints (\ref{eq:constraint1}) and
(\ref{eq:constraint2}). The covariant derivative is defined by
$(D\phi)_a^{~i}=\partial\phi_a^{~i}-iv_a^{~b}\phi_b^{~i}$. The
energy is bounded by tension
\begin{eqnarray}
T&=&\int dx \partial(\phi_a^{~i}M_i^{~j}\bar{\phi}_j^{~a}).\label{eq:tension}
\end{eqnarray}
The energy is saturated when the (anti-)BPS equations are satisfied
\begin{eqnarray}
(D\phi)_a^{~i}\mp(\phi_a^{~j}M_j^{~i}-\Sigma_a^{~b}\phi_b^{~i})=0.
\label{eq:bps_eq1}
\end{eqnarray}
We choose the upper sign so the BPS equation becomes
\begin{eqnarray}
(D\phi)_a^{~i}-(\phi_a^{~j}M_j^{~i}-\Sigma_a^{~b}\phi_b^{~i})=0.
\label{eq:bps_eq2}
\end{eqnarray}
We introduce complex matrix functions $S_a^{~b}(x)$ and
$f_a^{~i}(x)$ defined by
\begin{eqnarray}
\Sigma_a^{~b}-iv_a^{~b}=(S^{-1}\partial S)_a^{~b},~~~
\phi_a^{~i}=(S^{-1})_a^{~b}f_b^{~i}. \label{eq:field_redef1}
\end{eqnarray}
Then the BPS equation (\ref{eq:bps_eq2}) can be rewritten as
\begin{eqnarray}
\partial f_a^{~i}=f_a^{~j}M_j^{~i},
\end{eqnarray}
of which the solution is
\begin{eqnarray}
f_a^{~i}=H_{0a}^{~~j}(e^{Mx})_j^{~i}, \label{eq:field_redef2}
\end{eqnarray}
where $H_0$ is a complex constant matrix. As this matrix involves
the information of vacua and positions of domain walls, it is called
the moduli matrix \cite{Isozumi:2004jc,Isozumi:2004va}.

The BPS solution to (\ref{eq:bps_eq2}), obtained by combining (\ref{eq:field_redef1})
and (\ref{eq:field_redef2}) is
\begin{eqnarray}
 \phi_a^{~i}=(S^{-1})_a^{~b}H_{0b}^{~~j}(e^{Mx})_j^{~i}.\label{eq:phi_h}
\end{eqnarray}
From the definitions (\ref{eq:field_redef1}), $\Sigma$, $v$ and $\phi$ are invariant
under the transformation
\begin{eqnarray}
S_a^{\prime~b}=V_a^{~c}S_c^{~b},~~~
H_{0a}^{\prime~\,i}=V_a^{~c}H_{0c}^{~\,i}, \label{WVS}
\end{eqnarray}
where $V\in GL(N,\mathbf{C})$. The $V$ defines an equivalent class
of the sets of the matrix functions and moduli matrices $(S,H_{0})$.
This is called the world-volume symmetry
\cite{Isozumi:2004jc,Isozumi:2004va}.

Substituting (\ref{eq:phi_h}) into the constraints
(\ref{eq:constraint1}) and (\ref{eq:constraint2}), they become
\begin{eqnarray}
&&H_{0a}^{\,~i}(e^{2Mx})_i^{~j}{H_0^\dagger}_{j}^{~b}=(S\bar{S})_a^{~b}\equiv\Omega_a^{~b},
\label{eq:moduli_constraints1} \\
&&H_{0a}^{~i}J_{ij}H^{\mathrm{T}j}_{\,~b}=0.
\label{eq:moduli_constraints2}
\end{eqnarray}
The constraint (\ref{eq:moduli_constraints2}) together with the
world-volume symmetry (\ref{WVS}) gives a definition of
$SO(2N)/U(N)$ and $Sp(N)/U(N)$. Therefore $H_0$ parameterizes these
manifolds.

In order to analyze the BPS equation, it is useful to consider the
gauge invariant quantities \cite{Isozumi:2004va}. One of the
quantities is the identity component of $\Sigma$
\begin{eqnarray}
\Sigma_0=\frac{1}{4}\mathrm{tr}(S^{-1}(\partial\Omega)\Omega^{-1}S)
=\frac{1}{4}\frac{\partial\det\Omega}{\det\Omega},
\label{eq:sigma0}
\end{eqnarray}
where
\begin{eqnarray}
 \Sigma_0={\rm tr}(\Sigma)=\Sigma_1+\Sigma_2+\cdots+\Sigma_N.
\end{eqnarray}
The other is the tension (\ref{eq:tension}) in terms of $\Omega$
\begin{eqnarray}
T=\int
dx\partial(\phi_a^{~i}M_i^{~j}\bar{\phi}_j^{~a})=\frac{1}{2}\mathrm{tr}\partial(\Omega^{-1}\partial\Omega)
=\frac{1}{2}\partial^2\ln\det\Omega. \label{eq:energy}
\end{eqnarray}
We will use these quantities when we analyze the BPS domain wall
solutions.
%
%
\section{Wall solution in $SO(2N)/U(N)$ model}
\setcounter{equation}{0}
In this section we construct explicit BPS domain wall solutions for
massive $SO(2N)/U(N)$.
\subsection{$N=2$ case}\label{sec:so4}
We have shown that there exist two discrete vacua in $SO(2N)/U(N)$
in Section \ref{sec;massiveKahler}. We choose the vacua $\la 1 \ra$
and $\la 4 \ra$ in (\ref{vacO4}) without loss of generality.
We can express these vacua by the moduli matrix $H_0$:
\begin{eqnarray}
&& H_{0\la 1 \ra}=\left(
  \begin{array}{cccc}
   1 & 0 & 0 & 0 \\
   0 & 0 & 1 & 0\\
  \end{array}
 \right),~
H_{0\la 4 \ra}=\left(
  \begin{array}{cccc}
   0 & 1 & 0 & 0 \\
   0 & 0 & 0 & 1\\
  \end{array}
 \right),
 \label{eq:so4sp2vac}
\end{eqnarray}
which are related by (\ref{eq:phi_h}). Since there are two discrete
vacua, there is only one wall interpolating them. It should be an
elementary wall.

Before studying wall solutions in $SO(2N)/U(N)$, we briefly review
the property of walls in the Grassmann manifold
\cite{Isozumi:2004va}. In the paper walls are constructed
algebraically from elementary walls. By definition, an elementary
wall connects two nearest vacua of the same color index changing the
flavor by one unit $i\leftarrow i+1$. An elementary wall carrying
tension $T_{\la i\leftarrow i+1 \ra}$ is defined by
\begin{eqnarray}
[cM,a_i]=c(m_i-m_{i+1})a_i=T_{\la i\leftarrow i+1
\ra}a_i,\label{eq:ele_def}
\end{eqnarray}
where $c$ is a constant, $M$ is the mass matrix and $a_i$ is an
$N_f\times N_f$ square matrix generating an elementary wall. $N_f$
is the number of the flavors. From the first equality the mass
matrix $M$ and the matrix $a_i$ can be interpreted as a Cartan
generator and a step operator respectively. The $a_i$ has a nonzero
component only in the $(i,i+1)$-th element, which is equal to a
unit. With the use of $a_i$, an elementary wall is defined by
$H_{0\la A\leftarrow B\ra}=H_{0\la A \ra}e^{a_i(r)}$ where
$a_i(r)\equiv e^ra_i(r\in\mathbf{C})$ and $\la A \ra$ and $\la B
\ra$ are the vacua in the flavor $i$ and $i+1$ respectively in the
same color. The $e^{a_i(r)}$ is called the elementary-wall operator
\cite{Isozumi:2004va}.

We now turn to the $SO(2N)/U(N)$ model. Elementary walls changing
the flavor by one unit for the same color cannot be defined
consistently on the $SO(2N)/U(N)$ manifold. The moduli matrices of
elementary walls following the definition above do not satisfy the
constraint (\ref{eq:moduli_constraints2}), which stems from the
$F$-term constraint (\ref{const}). In addition, as it can be seen
from the vacua (\ref{eq:so4sp2vac}) of $SO(4)/U(2)$, changing the
flavor in the same color by one unit does not lead to the other
vacuum.

We shall modify the formalism of \cite{Isozumi:2004va} slightly for
$SO(2N)/U(N)$ case. We introduce an additional element with an
opposite sign into step operators $a_i$. We can construct elementary
walls constrained by (\ref{eq:moduli_constraints2}). We also choose
$U(N)$ gauge appropriately to label the vacua (\ref{vac_sigma})
keeping $\Sigma$ in a diagonal form. This can be formulated in terms
of the moduli matrices under the world-volume symmetry (\ref{WVS}).

The vacuum $H_{0\la 4 \ra}$ in (\ref{eq:so4sp2vac}) can be
transformed as
\begin{eqnarray}
H_{0\la 4 \ra}\rightarrow
\left(\begin{array}{cc}
 0 & -1 \\
 1 & 0
\end{array}\right)
\left(
  \begin{array}{cccc}
   0 & 1 & 0 & 0 \\
   0 & 0 & 0 & 1\\
  \end{array}
 \right)=
\left(
  \begin{array}{cccc}
   0 & 0 & 0 & -1 \\
   0 & 1 & 0 & 0\\
  \end{array}
 \right).
\end{eqnarray}
The elementary wall connecting the vacua $\la 1 \ra$ and $\la 4 \ra$
is
\begin{eqnarray}
 &
 H_{0\la 1\leftarrow 4 \ra}=\left(
  \begin{array}{cccc}
   1 & 0 & 0 & -e^r \\
   0 & e^r & 1 & 0\\
  \end{array}
 \right),~~~~r\in {\bf C},~~~
 -\infty<{\rm Re}(r)<\infty. \label{eq:so4_elementary}
\end{eqnarray}
We substitute (\ref{eq:so4_elementary}) into
(\ref{eq:moduli_constraints1}) and obtain
\begin{eqnarray}
S=\left(
\begin{array}{cc}
 S_1 & 0   \\
 0   & S_2 \\
\end{array}
\right),~~
\begin{array}{c}
S_1=\sqrt{e^{2m_1x}+e^{-2m_2x+2{\mathrm{Re}}(r)}},\\
S_2=\sqrt{e^{-2m_1x+2{\mathrm{Re}}(r)}+e^{2m_2x}}.
\end{array}
\end{eqnarray}
From (\ref{eq:phi_h}) we have the following solution
\begin{eqnarray}
\phi=\left(
\begin{array}{cccc}
 S_1^{-1}e^{m_1x} & 0                    &  0               &  -S_1^{-1}e^{-m_2x+r}  \\
 0                & S_2^{-1}e^{-m_1x+r}  & S_2^{-1}e^{m_2x} & 0                      \\
\end{array}
\right).
\end{eqnarray}
This has expected boundaries at $x\rightarrow \pm\infty$. The $\phi$
in those limits are
\begin{eqnarray}
\phi(x\rightarrow+\infty)&=& \left(
\begin{array}{cccc}
 1 & 0 &  0 & 0 \\
 0 & 0 &  1 & 0  \\
\end{array}
\right),\nonumber\\
\phi(x\rightarrow-\infty)&=&\left(
\begin{array}{cccc}
 0 & 0 &  0 & -1 \\
 0 & 1 &  0 & 0  \\
\end{array}
\right).\label{eq:vac_so4_phi}
\end{eqnarray}
The limit $\phi(x\rightarrow-\infty)$ in (\ref{eq:vac_so4_phi}) is
related to $\Phi_{\la 4 \ra}$ in (\ref{vacO4}) by $U(N)$ gauge
transformation. It shows that the wall (\ref{eq:so4_elementary})
connects the vacua $\la 1 \ra$ and $\la 4 \ra$. This result is
expected, considering that $SO(4)/U(2)$ is isomorphic to
$\mathbf{C}P^1$. A massive NLSM on $\mathbf{C}P^1$ has two discrete
vacua and there exits only one wall, which should be an elementary
wall. The results of $SO(4)/U(2)$ and $\mathbf{C}P^1$ are
consistent.

We use elementary-wall operators to construct walls from the next
section.
\subsection{$N=3$ case} \label{sec:so6}
%
As shown in \ref{sec:app1}, there are two sets of vacua $(\la 1 \ra,
\la 4 \ra, \la 6 \ra, \la 7 \ra)$ and $(\la 2 \ra, \la 3 \ra, \la 5
\ra, \la 8 \ra)$ which are related by parity. We choose the former
set of the vacua without loss of generality. We consider the
world-volume symmetry to 
obtain appropriate boundaries of walls as it
is done in Section \ref{sec:so4}. The moduli matrices for the vacua
are
\begin{eqnarray}
H_{0\la 1 \ra}=\left(
  \begin{array}{cccccc}
   1 & 0 & 0 & 0 & 0 & 0 \\
   0 & 0 & 1 & 0 & 0 & 0 \\
   0 & 0 & 0 & 0 & 1 & 0 \\
  \end{array}
 \right),\hspace{0.65cm}&&
H_{0\la 4 \ra}=\left(
  \begin{array}{cccccc}
   1 & 0 & 0 & 0 & 0 & 0 \\
   0 & 0 & 0 & 0 & 0 & -1 \\
   0 & 0 & 0 & 1 & 0 & 0 \\
  \end{array}
 \right),\nonumber
\end{eqnarray}
\begin{eqnarray}
H_{0\la 6 \ra}=\left(
  \begin{array}{cccccc}
   0 & 0 & -1 & 0 & 0 & 0 \\
   0 & 0 & 0  & 0 & 0 & -1 \\
   0 & 1 & 0  & 0 & 0 & 0 \\
  \end{array}
 \right),&&
H_{0\la 7 \ra}=\left(
  \begin{array}{cccccc}
   0 & 0 & 0 & 0  & 1 & 0 \\
   0 & 0 & 0 & -1 & 0 & 0 \\
   0 & 1 & 0 & 0  & 0 & 0 \\
  \end{array}
 \right).
 \label{eq:so6vac}
\end{eqnarray}
There are three matrices which generate elementary walls
\begin{eqnarray}
\begin{array}{ccc}
a_{1}=\left(
  \begin{array}{cccccc}
   0 & 0 & 0 & 0 & 0 & 0 \\
   0 & 0 & 0 & 0 & 0 & 0 \\
   0 & 0 & 0 & 0 & 0 & -1 \\
   0 & 0 & 0 & 0 & 0 & 0 \\
   0 & 0 & 0 & 1 & 0 & 0 \\
   0 & 0 & 0 & 0 & 0 & 0 \\
  \end{array}
 \right),&&
a_{2}=\left(
  \begin{array}{cccccc}
   0 & 0 & -1 & 0 & 0 & 0 \\
   0 & 0 & 0  & 0 & 0 & 0 \\
   0 & 0 & 0  & 0 & 0 &0 \\
   0 & 1 & 0  & 0 & 0 & 0 \\
   0 & 0 & 0  & 0 & 0 & 0 \\
   0 & 0 & 0  & 0 & 0 & 0 \\
  \end{array}
 \right), \\ \vspace{-0.5cm}\\
a_{3}=\left(
  \begin{array}{cccccc}
   0 & 0 & 0 & 0 & 0  & 0 \\
   0 & 0 & 0 & 0 & 0  & 0 \\
   0 & 0 & 0 & 0 & -1 &0 \\
   0 & 0 & 0 & 0 & 0  & 0 \\
   0 & 0 & 0 & 0 & 0  & 0 \\
   0 & 0 & 0 & 1 & 0  & 0 \\
  \end{array}
 \right).&&
 \label{eq:so6_creation_op}
\end{array}
\end{eqnarray}
We define $a_i(r)\equiv e^ra_i$ to obtain the moduli matrices for
elementary walls
\begin{eqnarray}
&&H_{0\la 1 \leftarrow 4 \ra}=H_{0\la 1\ra}e^{a_1(r_1)}=\left(
  \begin{array}{cccccc}
   1 & 0 & 0 & 0 & 0 & 0 \\
   0 & 0 & 1 & 0 & 0 & -e^{r_1} \\
   0 & 0 & 0 & e^{r_1} & 1 & 0 \\
  \end{array}
 \right),\nonumber\\
&&H_{0\la 4 \leftarrow 6 \ra}=H_{0\la 4\ra}e^{a_2(r_1)}=\left(
  \begin{array}{cccccc}
   1 & 0       & -e^{r_1} & 0 & 0 & 0 \\
   0 & 0       & 0        & 0 & 0 & -1 \\
   0 & e^{r_1} & 0        & 1 & 0 & 0 \\
  \end{array}
 \right),
 \nonumber\\
&&H_{0\la 6 \leftarrow 7 \ra}=H_{0\la 6\ra}e^{a_3(r_1)}=\left(
  \begin{array}{cccccc}
   0 & 0 & -1 & 0 & e^{r_1} & 0 \\
   0 & 0 & 0 & -e^{r_1} & 0 & -1 \\
   0 & 1 & 0 & 0 & 0 & 0 \\
  \end{array}
 \right),
 \label{eq:so6_single_elementaryA}
\end{eqnarray}
where the parameters $r_i$ $(i=1,2,\cdots)$ are complex numbers
ranging $-\infty<\mathrm{Re}(r_i)<\infty$. We follow the convention
that the operators act on the moduli matrices from the right.

Non-vanishing commutation relations of matrices in
(\ref{eq:so6_creation_op}) define compressed single walls. A
compressed wall generated by compressing $n$ elementary walls is a
compressed wall of level $n-1$ \cite{Isozumi:2004va}. There are two
compressed walls of level one generated by
\begin{eqnarray}
 E_1=[a_1,a_2]\neq0,~~~E_2=[a_2,a_3]\neq0. \label{eq:so6_gen_comp1}
\end{eqnarray}

\begin{figure}[h!]
\begin{center}
\epsfxsize=7cm
   \epsfbox{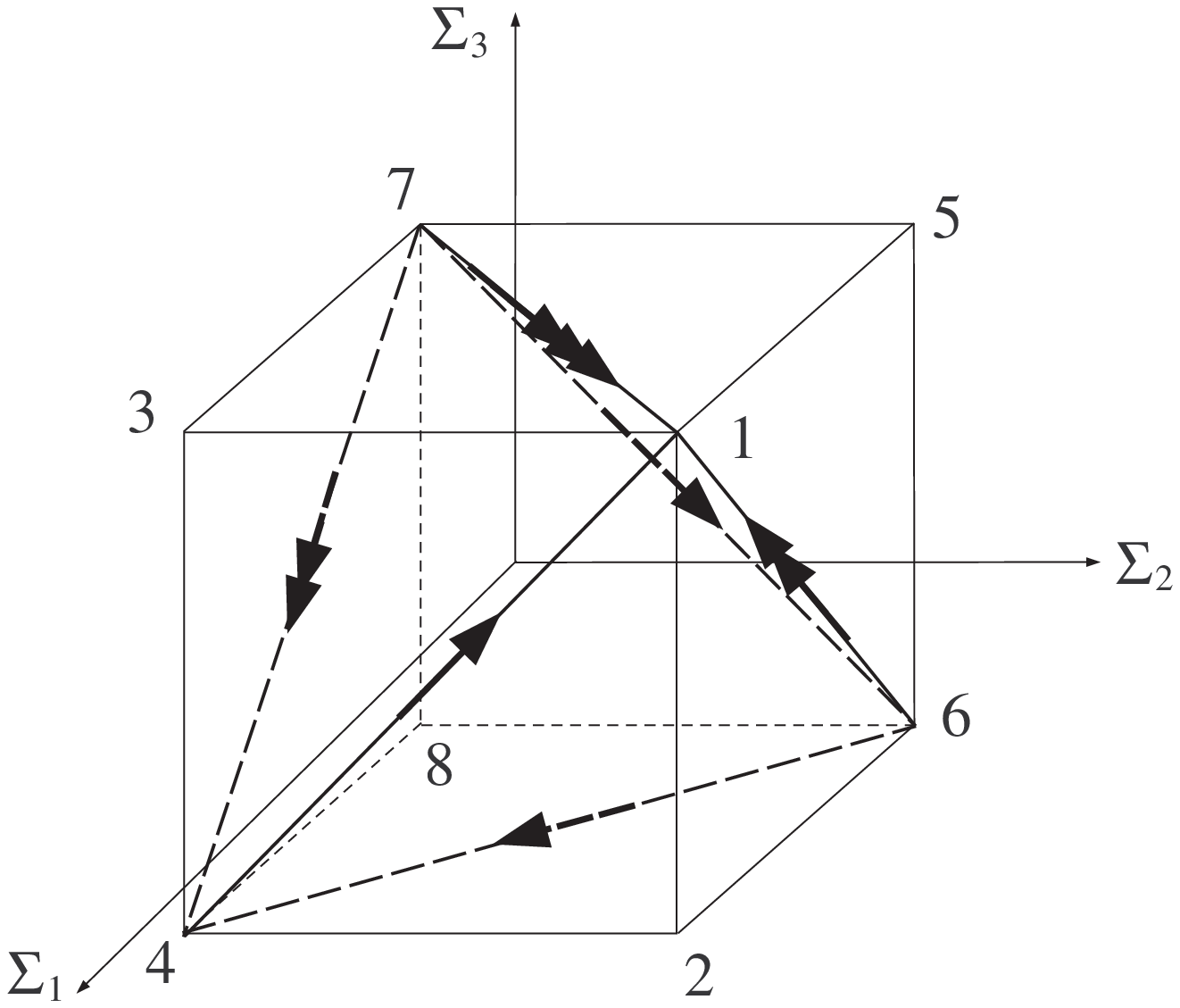}\\
   $\mathrm{(a)}$\\
$\begin{array}{cc}
  \epsfxsize=7cm
   \epsfbox{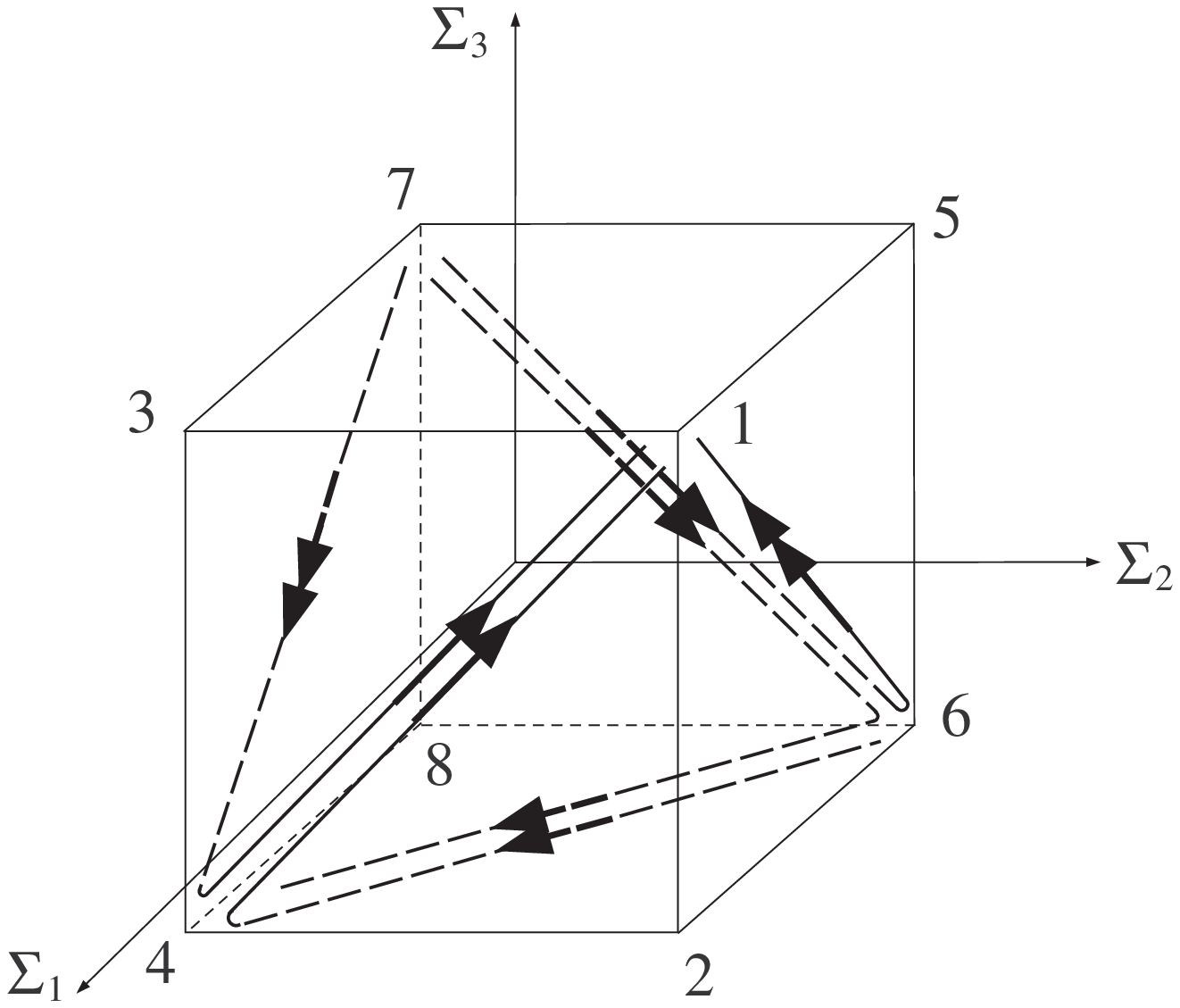}
  &
  \epsfxsize=7cm
   \epsfbox{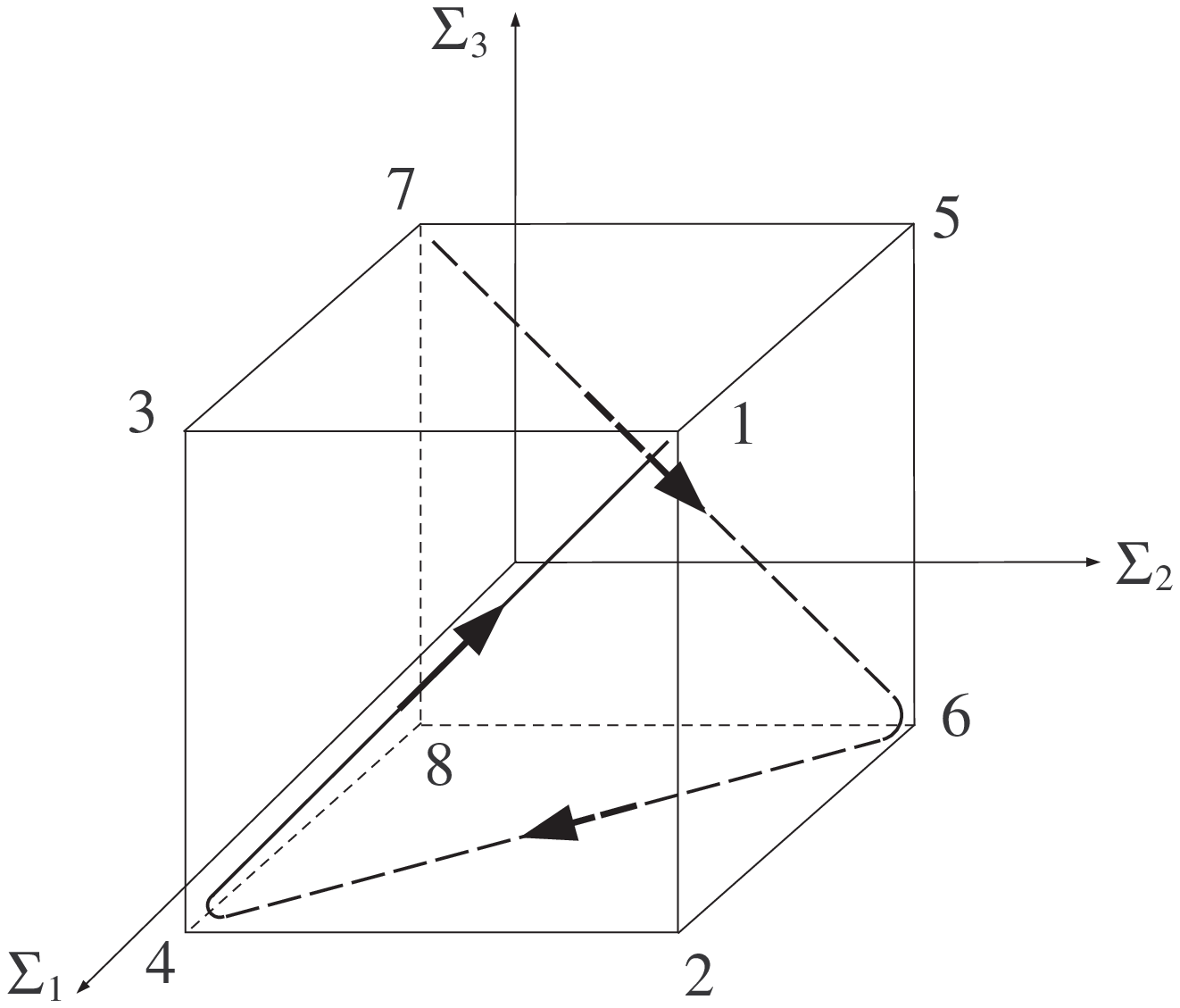}\\
   \mathrm{(b)}&\mathrm{(c)}
\end{array}$
  \caption{(a)Single walls. Arrows with one arrowhead denote elementary single walls and arrows with two(three)
arrowheads denote compressed single walls of level one(two).
(b)Double walls. (c)A triple wall.}
  \label{fig:so6}
\end{center}
\end{figure}

The matrix $E_1$ generates a wall, which interpolates the vacua $\la
1 \ra$ and $\la 6 \ra$ while the matrix $E_2$ generates a wall,
which interpolates the vacua $\la 4 \ra$ and $\la 7 \ra$:
\begin{eqnarray}
&&H_{0\la 1 \leftarrow 6 \ra}=H_{0\la 1\ra}e^{E_1(r_1)}=\left(
  \begin{array}{cccccc}
   1 & 0       & 0  & 0 & 0 & -e^{r_1} \\
   0 & 0       & 1  & 0 & 0 & 0        \\
   0 & e^{r_1} & 0  & 0 & 1 & 0        \\
  \end{array}
 \right), \label{eq:so6_single_compressedA1}\\
&&H_{0\la 4 \leftarrow 7 \ra}=H_{0\la 4\ra}e^{E_2(r_1)}=\left(
  \begin{array}{cccccc}
   1 & 0        & 0  & 0 & e^{r_1} & 0 \\
   0 & e^{r_1}  & 0  & 0 & 0       & -1 \\
   0 & 0        & 0  & 1 & 0       & 0 \\
  \end{array}
 \right),
 \label{eq:so6_single_compressedA2}
\end{eqnarray}
where $E_i(r)\equiv e^rE_i$. A compressed wall of level one and an
elementary wall can be compressed to be a compressed wall of level
two. The corresponding wall is generated by the operator $E_3$
\begin{eqnarray}
E_3=[a_1,E_2]=[E_1,a_3]\neq 0. \label{eq:so6_gen_comp2}
\end{eqnarray}
The first commutator describes compression of the elementary wall
$H_{0\la 1 \leftarrow 4 \ra}$ and the compressed wall of level one
$H_{0\la 4 \leftarrow 7 \ra}$. The second commutator describes
compression of the compressed wall of level one $H_{0\la 1\leftarrow
6 \ra}$ and the elementary wall $H_{0\la 6 \leftarrow 7 \ra}$. As it
can also be seen from Figure \ref{fig:so6}(a), the both of them
leads to the same wall interpolating the vacua $\la 1 \ra$ and $\la
7 \ra$
\begin{eqnarray}
H_{0\la 1 \leftarrow 7 \ra}=H_{0\la 1\ra}e^{E_3(r_1)}=\left(
  \begin{array}{cccccc}
   1 & 0        & 0  &  -e^{r_1} & 0  & 0 \\
   0 & e^{r_1}  & 1  &  0        & 0  & 0 \\
   0 & 0        & 0  &  0        & 1  & 0 \\
  \end{array}
 \right).
\end{eqnarray}
Configurations of the single walls in $\Sigma$-space are drawn in
Figure \ref{fig:so6}(a).

\begin{figure}[t]
\begin{center}
 $\begin{array}{ccc}
  \epsfxsize=5cm
   \epsfbox{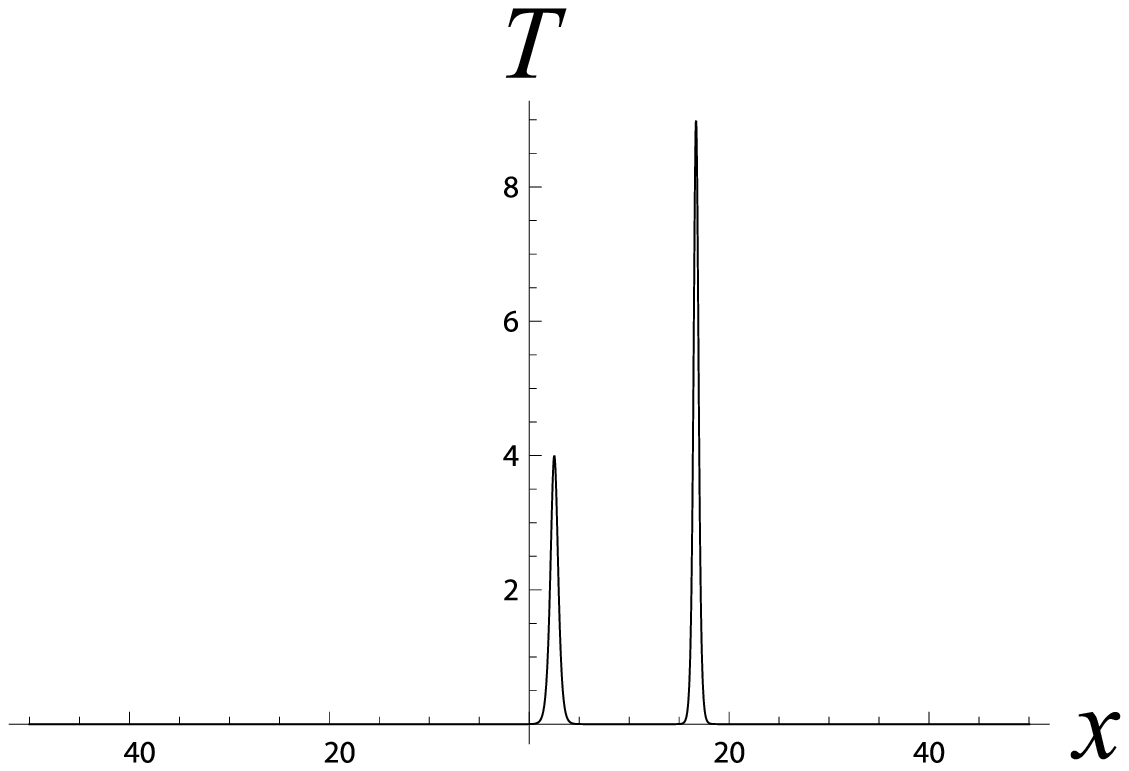}
  &
  \epsfxsize=5cm
   \epsfbox{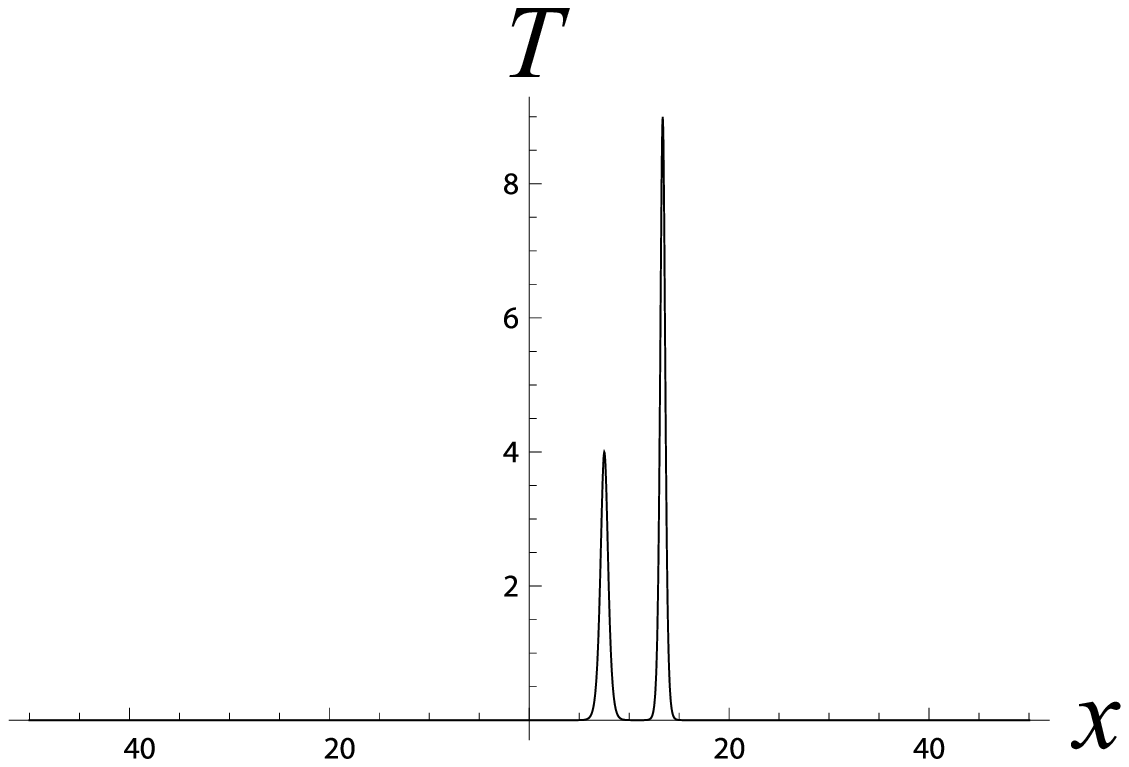}
  &
  \epsfxsize=5cm
   \epsfbox{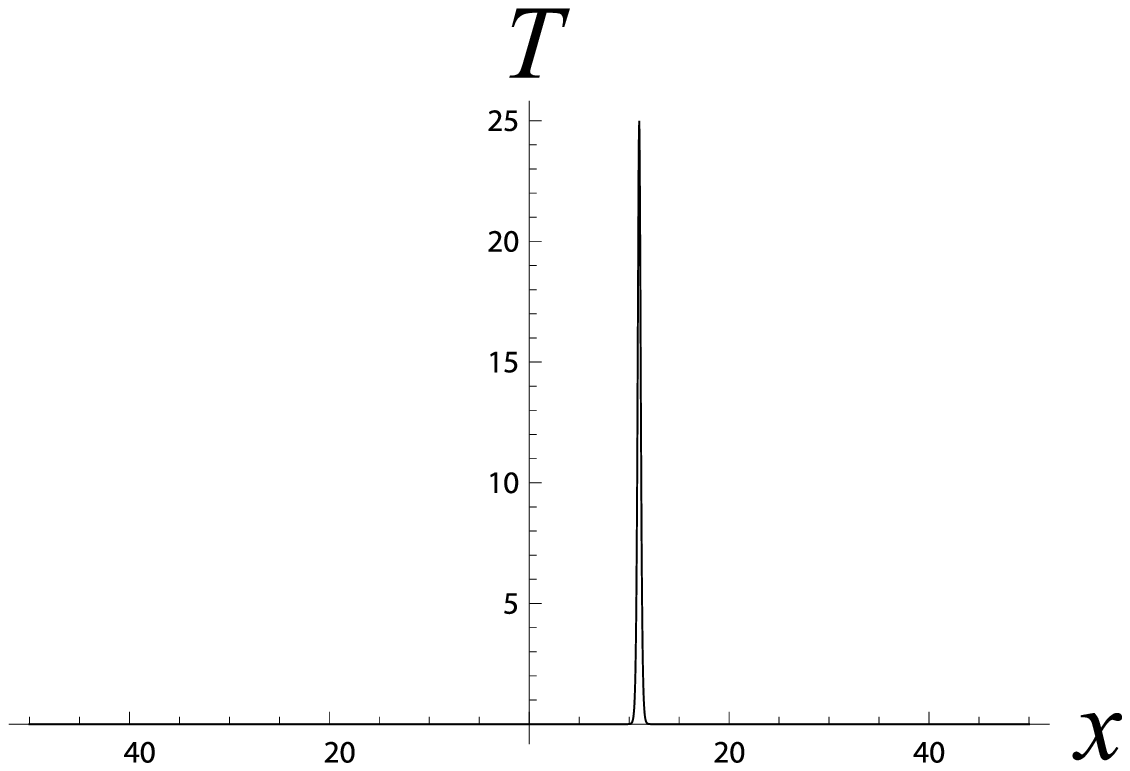}
  \\~\\
  \epsfxsize=5cm
   \epsfbox{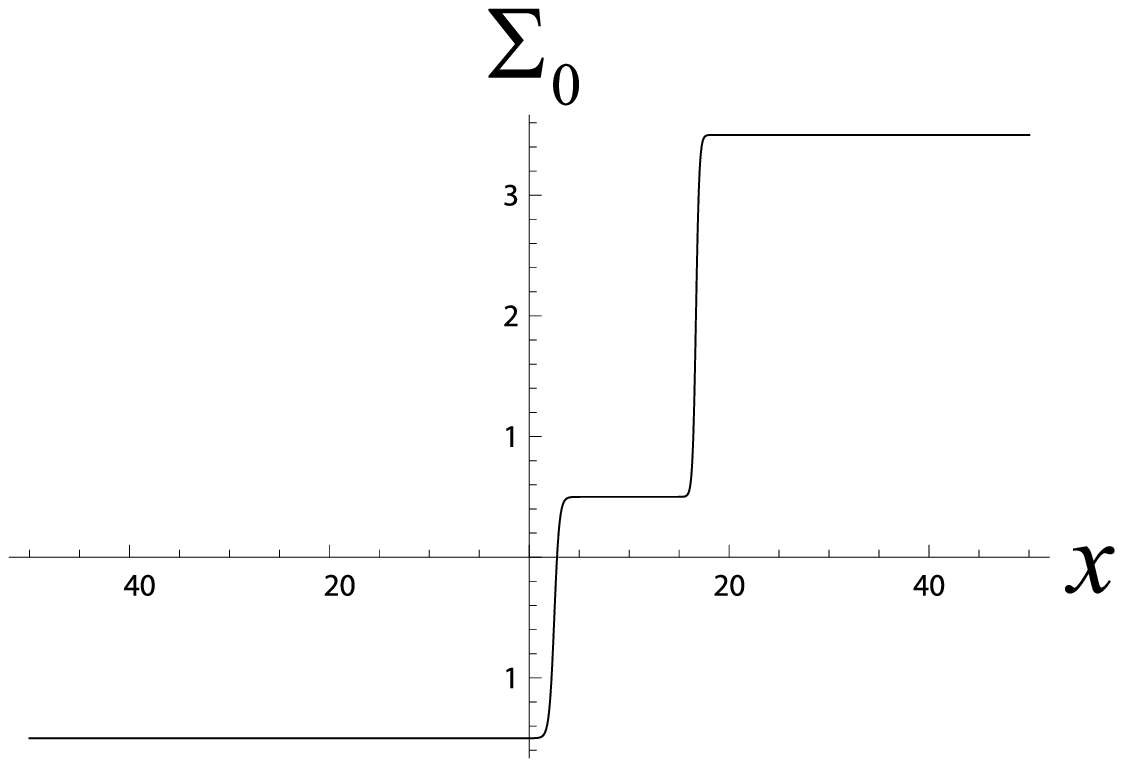}
  &
  \epsfxsize=5cm
   \epsfbox{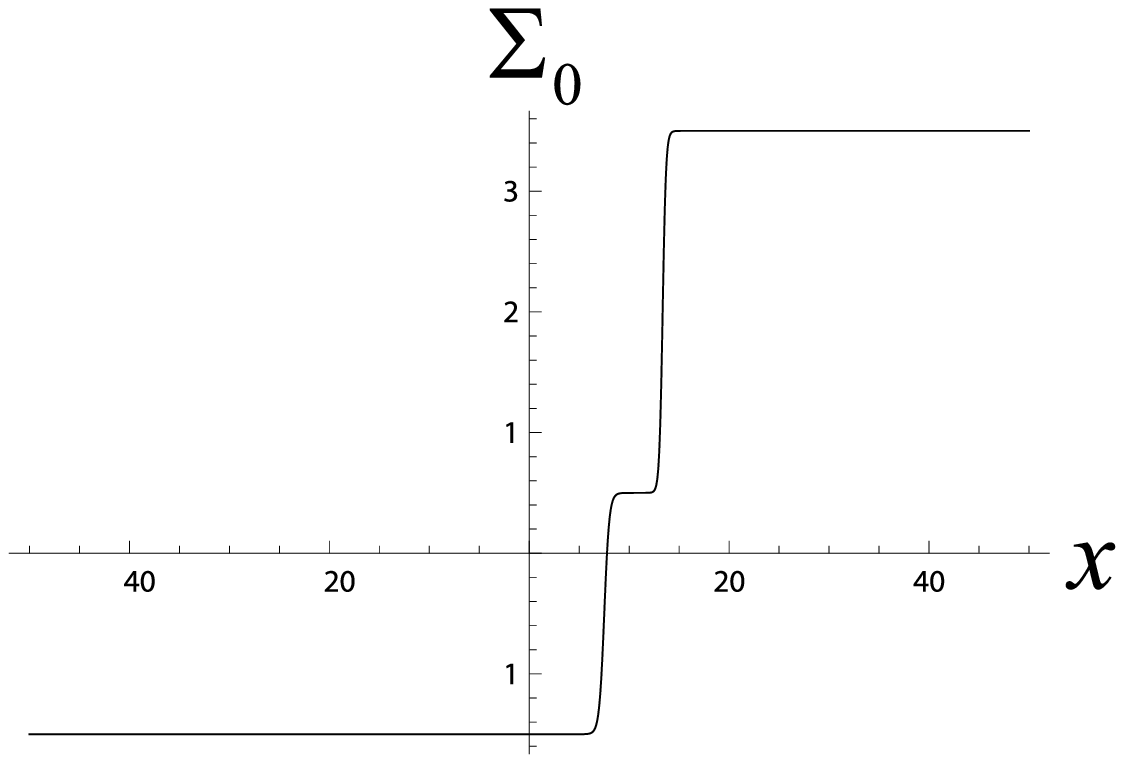}
  &
  \epsfxsize=5cm
   \epsfbox{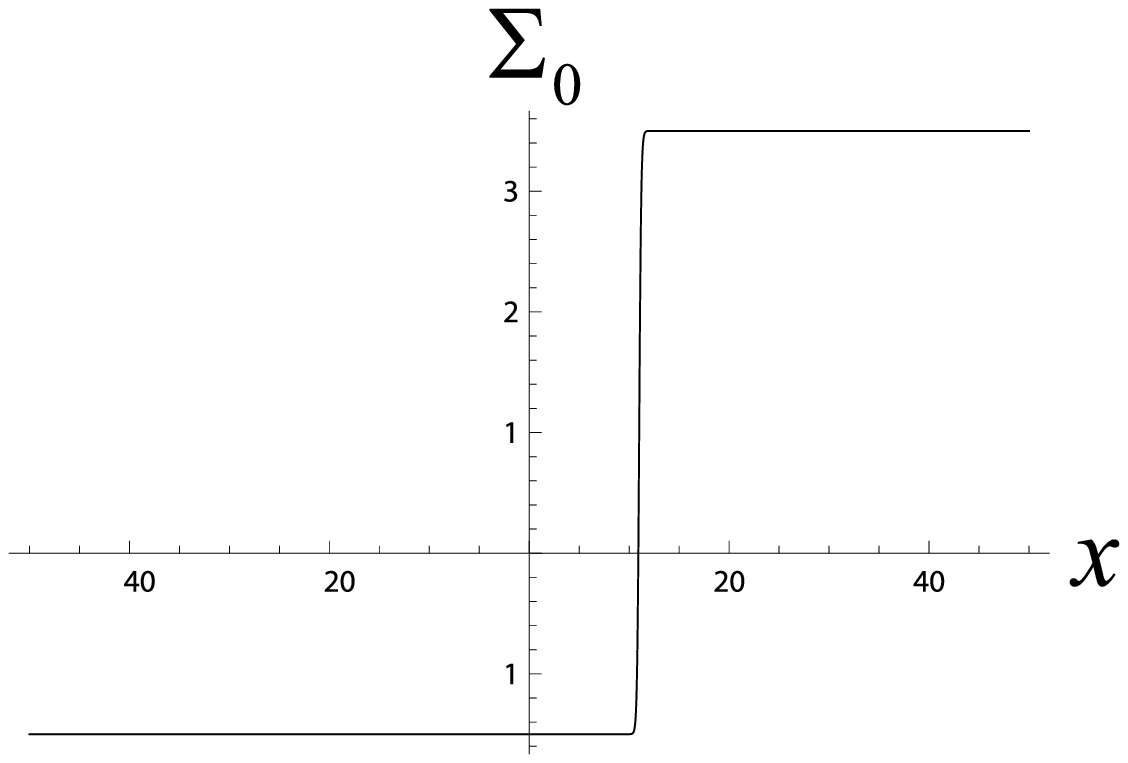}
  \end{array}$
  \caption{Plots of energy density $T$ and gauge component $\Sigma_0$ with masses $m_1=4$, $m_2=2$, $m_3=1$.
  Two elementary walls forming a double wall become a compressed wall
  of level two as
  $r_1$ and $r_2$ vary.
  $(r_1,r_2)=(50,5)$ in the
  left, $(r_1,r_2)=(40,15)$ in the middle and $(r_1,r_2)=(-10,65)$ in the right.}
  \label{fig:so6_double_to_compressed}
\end{center}
\end{figure}

Next we consider multiwall solutions. A double wall is constructed
by multiplying an elementary-wall operator to the moduli matrix of a
single wall. There are four double walls:
\begin{eqnarray}
&&H_{0\la 1 \leftarrow 4 \leftarrow 6\ra}=H_{0\la 1 \leftarrow 4
\ra}e^{a_2(r_2)}=\left(
  \begin{array}{cccccc}
   1 & 0           & -e^{r_2} & 0       & 0 & 0        \\
   0 & 0           & 1        & 0       & 0 & -e^{r_1} \\
   0 & e^{r_1+r_2} & 0        & e^{r_1} & 1 & 0        \\
   \end{array}
  \right),\\
&&H_{0\la 4 \leftarrow 6 \leftarrow 7\ra}=H_{0\la 4 \leftarrow 6
\ra}e^{a_3(r_2)}=\left(
  \begin{array}{cccccc}
   1 & 0       & -e^{r_1} & 0         & e^{r_1+r_2} & 0 \\
   0 & 0       & 0        & -e^{r_2}  & 0           & -1 \\
   0 & e^{r_1} & 0        & 1         & 0           & 0 \\
  \end{array}
 \right), \\
&&H_{0\la 1 \leftarrow 4 \leftarrow 7\ra}=H_{0\la 1 \leftarrow 4
\ra}e^{E_2(r_2)}=\left(
  \begin{array}{cccccc}
   1 & 0           & 0 & 0       & e^{r_2} & 0        \\
   0 & e^{r_1+r_2} & 1 & 0       & 0       & -e^{r_1} \\
   0 & 0           & 0 & e^{r_1} & 1       & 0        \\
  \end{array}
 \right),\\
&&H_{0\la 1 \leftarrow 6 \leftarrow 7\ra}=H_{0\la 1 \leftarrow 6
\ra}e^{a_3(r_2)}=\left(
  \begin{array}{cccccc}
   1 & 0       & 0 & -e^{r_1+r_2} & 0        & -e^{r_1} \\
   0 & 0       & 1 & 0            & -e^{r_2} & 0        \\
   0 & e^{r_1} & 0 & 0            & 1        & 0        \\
  \end{array}
 \right).
\end{eqnarray}
For instance, $\la 1\leftarrow 4 \leftarrow 6 \ra$ means
configuration interpolating three vacua, $\la 1 \ra$ at $x=+\infty$,
$\la 4 \ra$ at $-\infty<x<\infty$ and $\la 6 \ra$ at $x=-\infty$.
The double walls $H_{0\la 1 \leftarrow 4 \leftarrow 6\ra}$ and
$H_{0\la 4 \leftarrow 6 \leftarrow 7\ra}$ are composed of two
elementary walls. The double walls $H_{0\la 1 \leftarrow 4
\leftarrow 7\ra}$ and $H_{0\la 1 \leftarrow 6 \leftarrow 7\ra}$ are
composed of one elementary wall and one compressed wall of level one
as it is described in Figure \ref{fig:so6}(b).
We observe compression of two elementary walls from the double wall
$H_{0\la 1 \leftarrow 4 \leftarrow 6\ra}$ as an example. Under the
world-volume symmetry, the moduli matrix transforms
\begin{eqnarray}
H_{0\la 1 \leftarrow 4 \leftarrow 6\ra}
&\rightarrow&
\left(\begin{array}{ccc}
   1 & e^{r_2} & 0  \\
   0 & 1       & 0  \\
   0 & 0       & 1  \\
  \end{array}\right)
\left(
  \begin{array}{cccccc}
   1 & 0           & -e^{r_2} & 0       & 0 & 0        \\
   0 & 0           & 1        & 0       & 0 & -e^{r_1} \\
   0 & e^{r_1+r_2} & 0        & e^{r_1} & 1 & 0        \\
   \end{array}
  \right)\nonumber\\
&=&\left(
  \begin{array}{cccccc}
   1 & 0           & 0  & 0       & 0  & -e^{r_1+r_2} \\
   0 & 0           & 1  & 0       & 0  & -e^{r_1}     \\
   0 & e^{r_1+r_2} & 0  & e^{r_1} & 1  & 0            \\
   \end{array}
  \right). \label{SO6-limit}
\end{eqnarray}
Keeping the parameter $r_1+r_2$ finite in the limit of
$r_1\rightarrow-\infty$, it leads to a compressed wall of level one,
which interpolates the vacua $\la 1 \ra$ and $\la 6 \ra$. This is
illustrated in Figure \ref{fig:so6_double_to_compressed}. The upper
panels in Figure \ref{fig:so6_double_to_compressed} show the energy
density of the multiwall configuration $H_{0\la 1\leftarrow 4
\leftarrow 6\ra}$. As $r_1-r_2$ becomes large, two walls approach
and get compressed to a single wall. Such a compression can be also
seen in the lower panels in Figure
\ref{fig:so6_double_to_compressed}. They show the configuration of
$\Sigma_0$ defined in (\ref{eq:sigma0}) which displays two kinks
with the energy density according to $r_1$ and $r_2$.

Finally we consider a triple wall. A triple wall is obtained by
multiplying an elementary-wall operator to a moduli matrix for a
double wall. There is only one triple wall
\begin{eqnarray}
H_{0\la 1 \leftarrow 4 \leftarrow 6  \leftarrow 7\ra}=H_{0\la 1
\leftarrow 4 \leftarrow 6 \ra}e^{a_3(r_3)}=\left(
  \begin{array}{cccccc}
   1 & 0           & -e^{r_2} & 0            & e^{r_2+r_3} & 0        \\
   0 & 0           & 1        & -e^{r_1+r_3} & -e^{r_3}    & -e^{r_1} \\
   0 & e^{r_1+r_2} & 0        & e^{r_1}      & 1           & 0        \\
  \end{array}
 \right),
\end{eqnarray}
which is composed of three elementary walls as described in Figure
\ref{fig:so6}(c).
This is the multiwall composed of the maximal number of single walls in $SO(6)/U(3)$.
In $SO(6)/U(3)$, we have found that there are six single walls,
consisting of three elementary walls, two compressed wall of level
one and two compressed wall of level two, four double walls and one
triple wall. We compare this result with $\mathbf{C}P^3$, which is
isomorphic to $SO(6)/U(3)$ in \ref{sec:app2}.
%
%
%
\section{Wall solution in $Sp(N)/U(N)$ model}
\setcounter{equation}{0}
In this section we construct explicit BPS domain wall solutions for
massive $Sp(N)/U(N)$.
\subsection{$N=1$ case}
There are two vacua in $Sp(1)/U(1)$
\begin{eqnarray}
&&\Phi_{\la 1 \ra}=(1,0),~~\Sigma=m, \nonumber\\
&&\Phi_{\la 2 \ra}=(0,\alpha),~~\Sigma=-m,~~(\alpha=\pm1),
\end{eqnarray}
from the vacuum condition (\ref{eq:vac_equation1}) and
(\ref{eq:vac_equation2}). The corresponding moduli matrices are
\begin{eqnarray}
H_{0\la 1 \ra}=(1,0),~~H_{0\la 2 \ra}=(0,1).
\end{eqnarray}
We have removed the plus-minus sign of $\alpha$ by the world-volume
symmetry (\ref{WVS}). There is only one single wall therefore which
is an elementary wall interpolating the vacua
\begin{eqnarray} H_{0\la 1 \leftarrow 2 \ra}
=(1,e^r).\label{eq:sp1_single_elementary}
\end{eqnarray}
This result is consistent with $\mathbf{C}P^1$, which is isomorphic
to $Sp(1)/U(1)$.
%
\subsection{$N=2$ case}\label{sec:sp2}
\begin{figure}[t]
\begin{center}
 $\begin{array}{cc}
  \epsfxsize=7cm
   \epsfbox{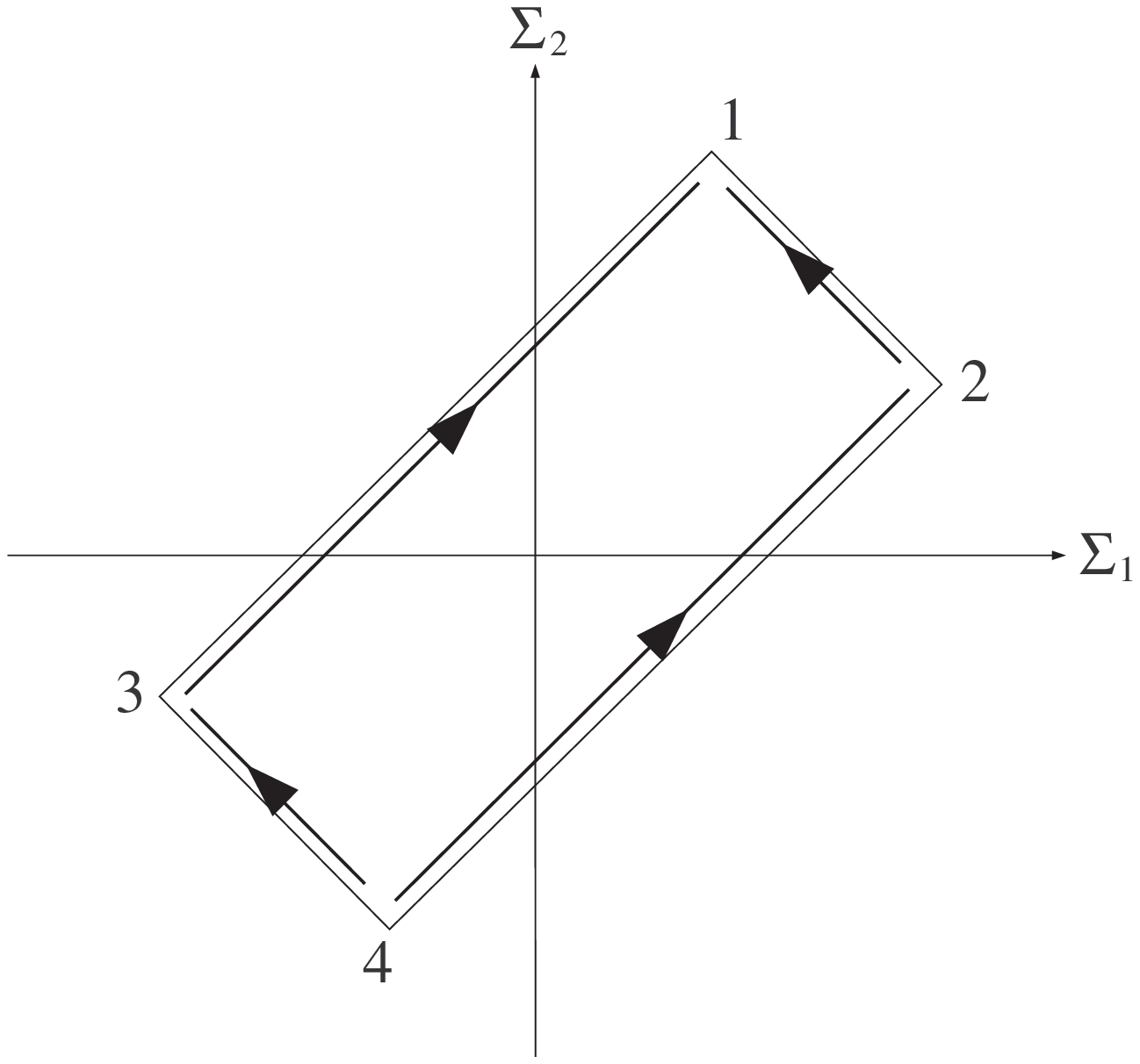}
   &
  \epsfxsize=7cm
   \epsfbox{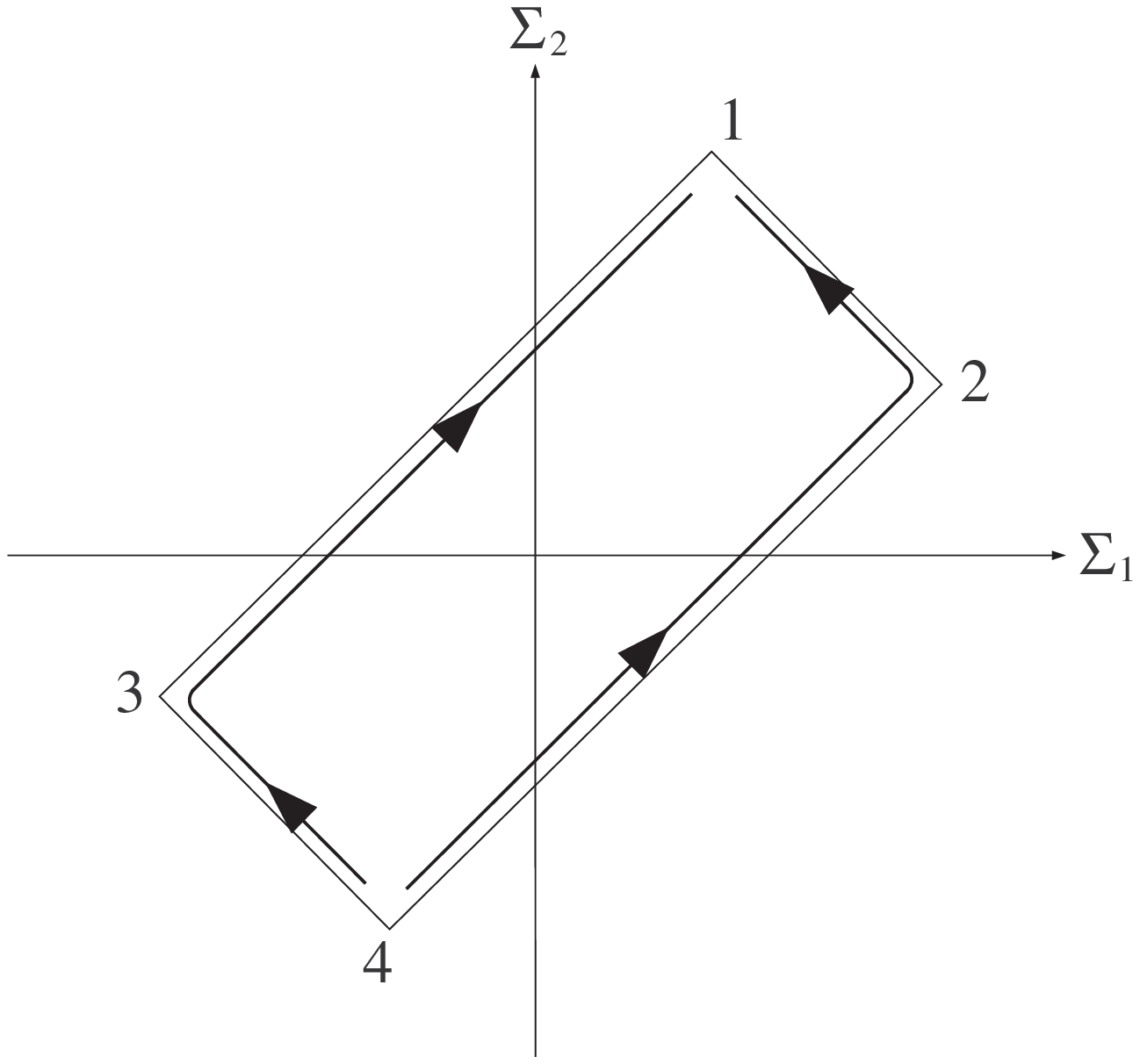}\\
  \mathrm{(a)} & \mathrm{(b)}\\
  \end{array}$
  \caption{Walls in $Sp(2)/U(2)$. (a) Single walls. (b)Double walls, which are penetrable.}
  \label{fig:sp2_single_double}
\end{center}
\end{figure}

The vacua of $Sp(2)/U(2)$ are (\ref{vacO4}). The moduli matrices for
the vacua related by (\ref{eq:phi_h}) are
\begin{eqnarray}
&&H_{0\la 1 \ra}=\left(
  \begin{array}{cccc}
   1 & 0 & 0 & 0 \\
   0 & 0 & 1 & 0\\
  \end{array}
 \right),~
 H_{0\la 2 \ra}=\left(
 \begin{array}{cccc}
  1 & 0 & 0 & 0 \\
  0 & 0 & 0 & 1\\
 \end{array}\right),
 \nonumber\\
&&H_{0\la 3 \ra}=\left(
 \begin{array}{cccc}
  0 & 1 & 0 & 0 \\
  0 & 0 & 1 & 0\\
 \end{array}
\right),~
H_{0\la 4 \ra}=\left(
  \begin{array}{cccc}
   0 & 1 & 0 & 0 \\
   0 & 0 & 0 & 1\\
  \end{array}
 \right).
\end{eqnarray}
The plus-minus sign of $\alpha_i$ for $Sp(2)/U(2)$ in (\ref{vacO4})
has been removed by the world-volume symmetry (\ref{WVS}).
Elementary walls interpolating the four vacua are generated by two
operators
\begin{eqnarray}
a_1=\left(
  \begin{array}{cccc}
   0 & 0 & 0 & 0 \\
   0 & 0 & 0 & 0 \\
   0 & 0 & 0 & 1 \\
   0 & 0 & 0 & 0 \\
  \end{array}
 \right),~~
a_2=\left(\begin{array}{cccc}
   0 & 1 & 0 & 0 \\
   0 & 0 & 0 & 0 \\
   0 & 0 & 0 & 0 \\
   0 & 0 & 0 & 0 \\
  \end{array}
 \right).
\label{eq:sp2_generators}
\end{eqnarray}
There are twelve elementary single walls
\begin{eqnarray}
H_{0\la 1 \leftarrow 2\ra}&=&H_{0\la 1\ra}e^{a_1(r_1)}=\left(
  \begin{array}{cccc}
   1 & 0 & 0 & 0 \\
   0 & 0 & 1 & e^{r_1}\\
  \end{array}
 \right),\nonumber\\
H_{0\la 3 \leftarrow 4\ra}&=&H_{0\la 3\ra}e^{a_1(r_1)}=\left(
  \begin{array}{cccc}
   0 & 1 & 0 & 0 \\
   0 & 0 & 1 & e^{r_1}\\
  \end{array}
 \right),\nonumber\\
H_{0\la 1 \leftarrow 3\ra}&=&H_{0\la 1\ra}e^{a_2(r_1)}=\left(
  \begin{array}{cccc}
   1 & e^{r_1} & 0 & 0 \\
   0 & 0 & 1 & 0\\
  \end{array}
 \right),\nonumber\\
H_{0\la 2 \leftarrow 4\ra}&=&H_{0\la 2\ra}e^{a_2(r_1)}=\left(
  \begin{array}{cccc}
   1 & e^{r_1} & 0 & 0 \\
   0 & 0 & 0 & 1\\
  \end{array}
 \right),
\end{eqnarray}
where $a_i(r)\equiv e^ra_i$. The walls are drawn in Figure
\ref{fig:sp2_single_double}(a).
The operators in (\ref{eq:sp2_generators}) commute:
\begin{eqnarray}
[a_1,a_2]=0.\label{eq:sp2_commutator}
\end{eqnarray}
Therefore no compressed wall exists in $Sp(2)/U(2)$.

\begin{figure}[t]
\begin{center}
 $\begin{array}{ccc}
  \epsfxsize=5cm
   \epsfbox{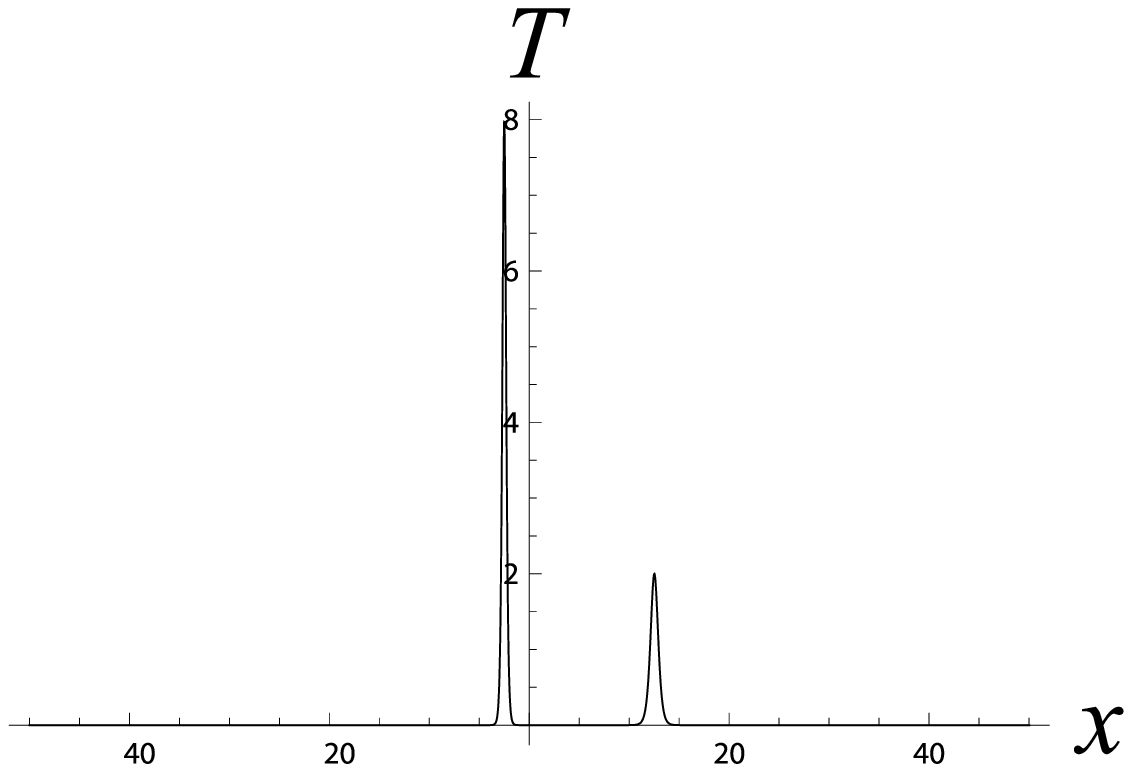}
  &
  \epsfxsize=5cm
   \epsfbox{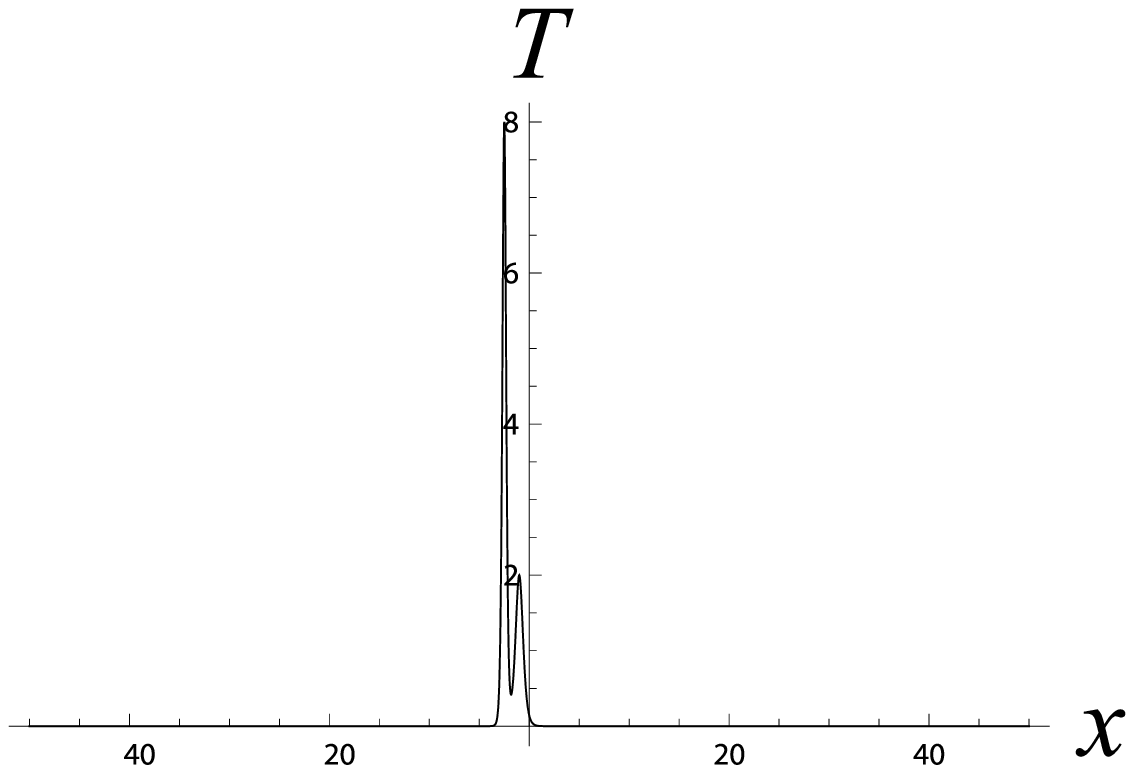}
  &
  \epsfxsize=5cm
   \epsfbox{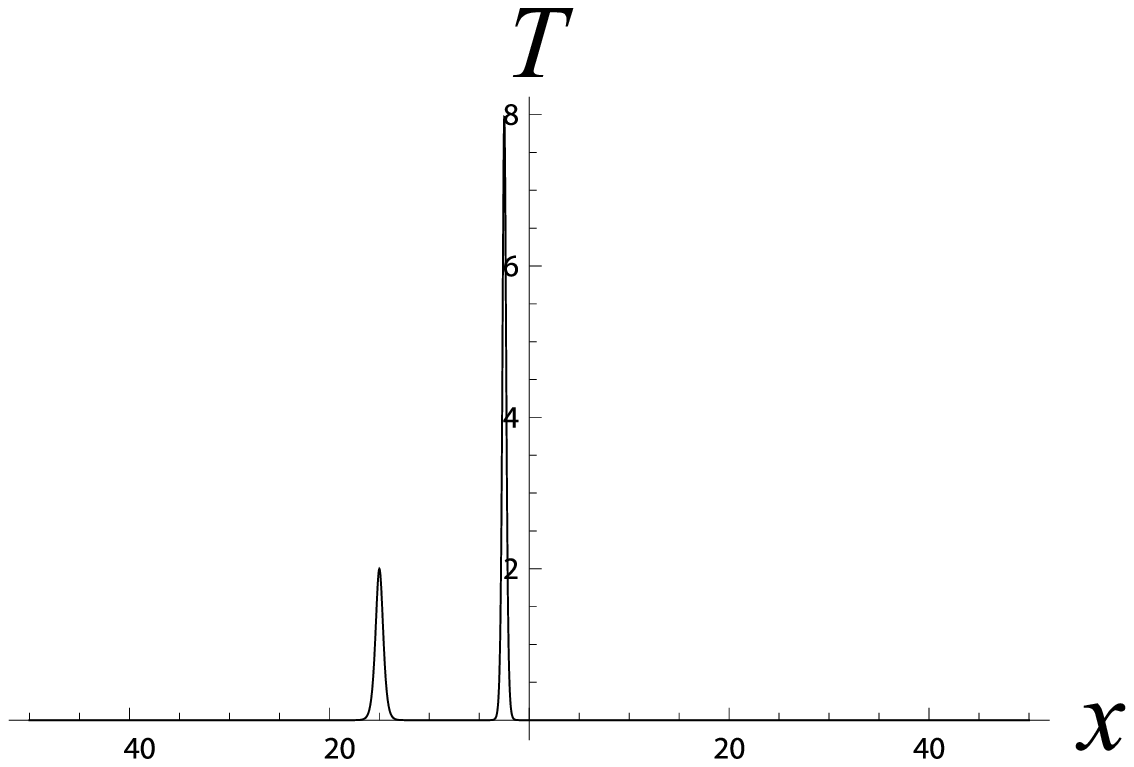}
  \\~\\
  \epsfxsize=5cm
   \epsfbox{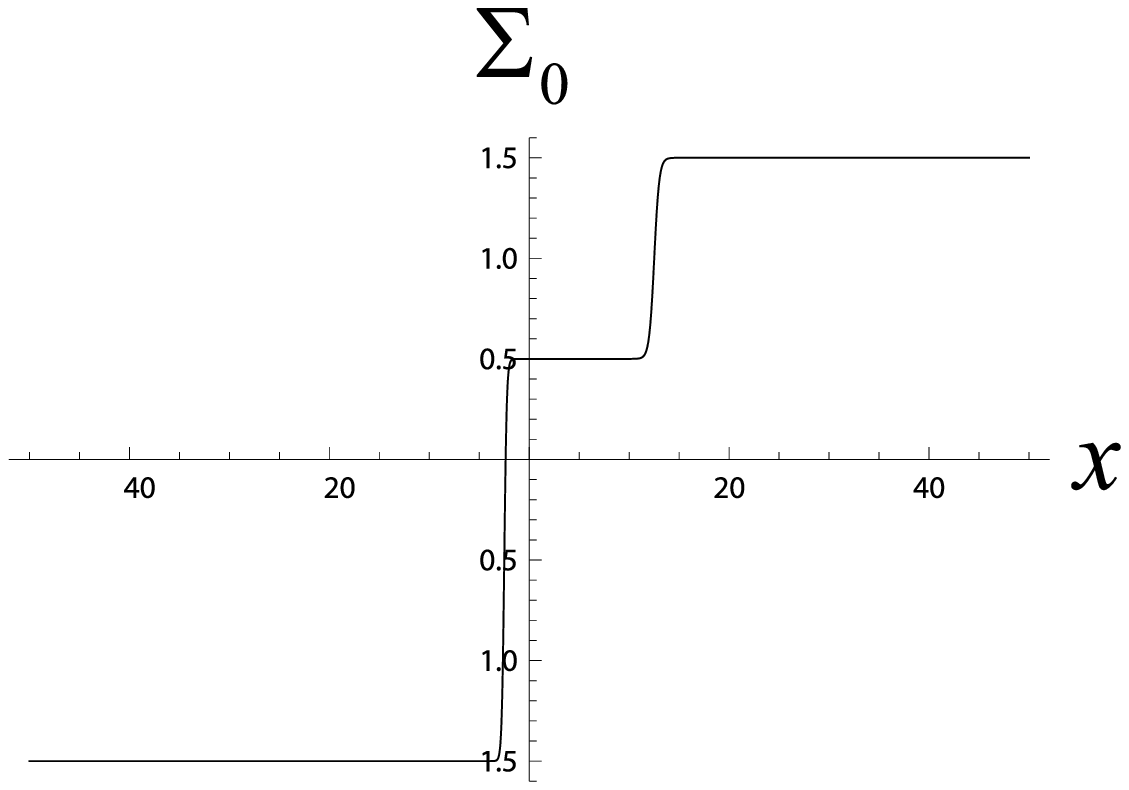}
  &
  \epsfxsize=5cm
   \epsfbox{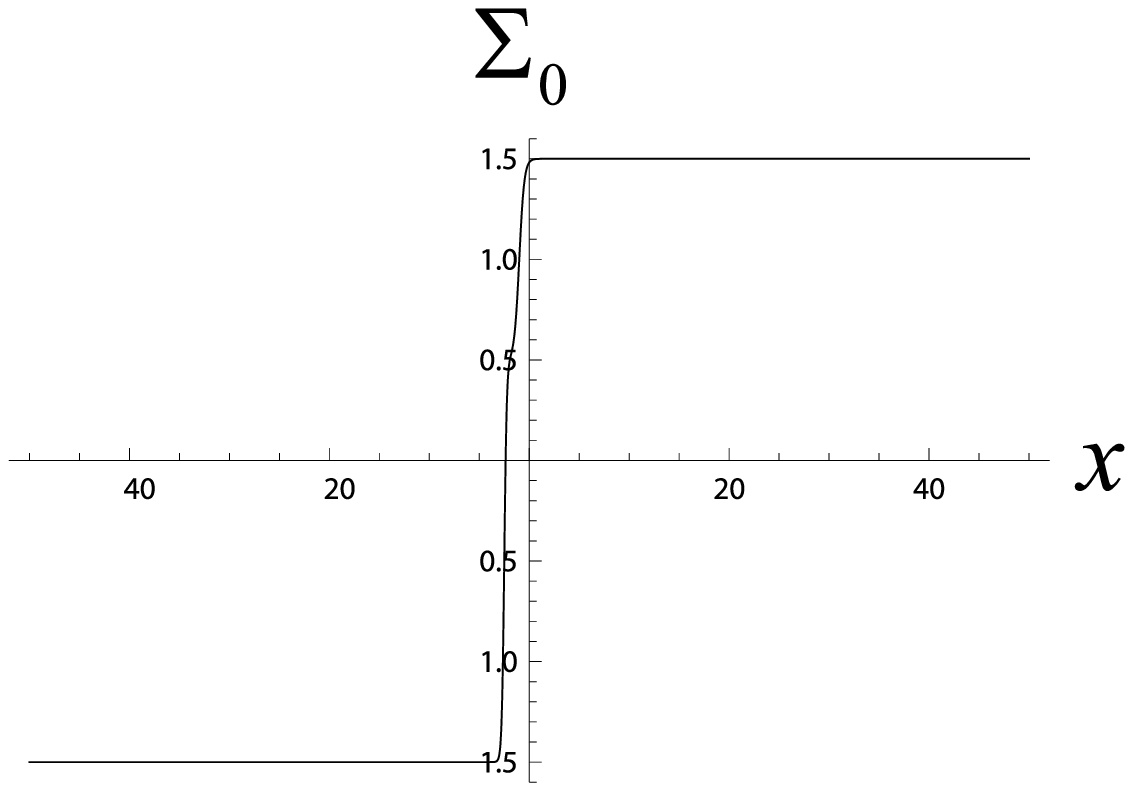}
  &
  \epsfxsize=5cm
   \epsfbox{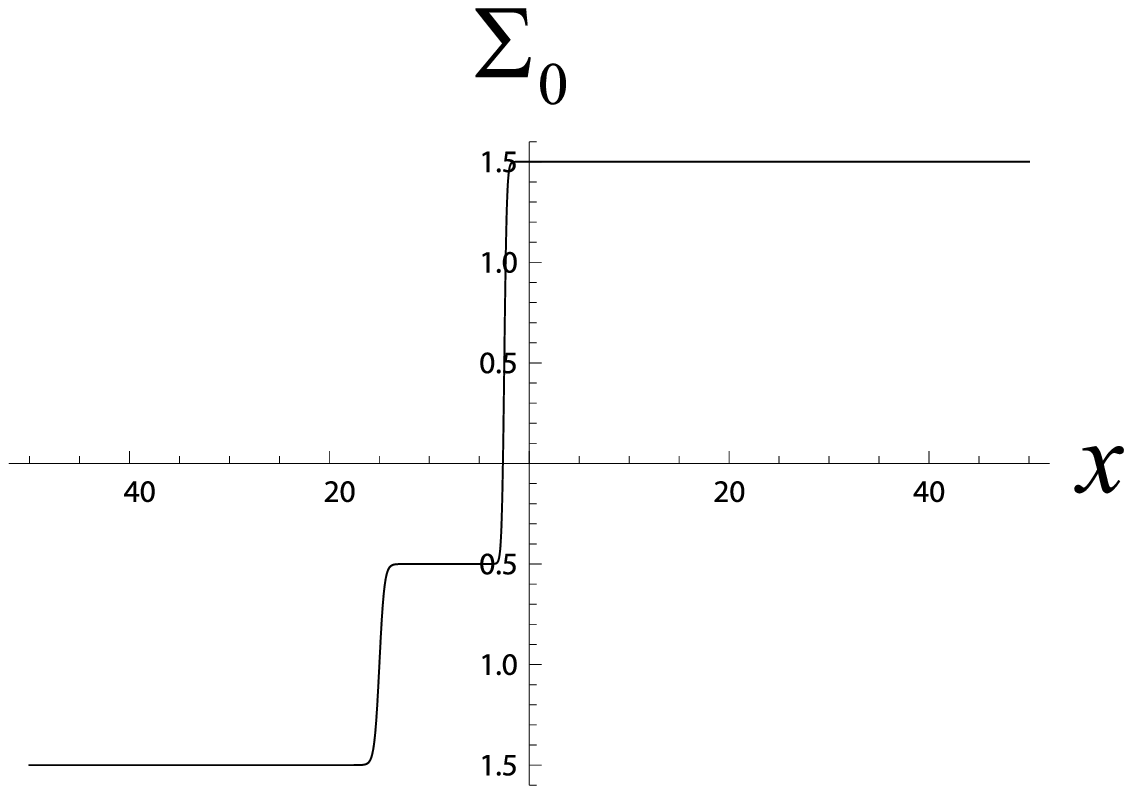}
  \end{array}$
  \caption{Plots of energy density $T$ and gauge component $\Sigma_0$ with $m_1=2$, $m_2=1$.
  Two double walls are penetrable. $(r_1,r_2)=(25,-10)$ in the left, $(r_1,r_2)=(-2,-10)$ in the middle
  and $(r_1,r_2)=(-30,-10)$ in the right.}
  \label{fig:sp2_double_penetrable}
\end{center}
\end{figure}
We construct double walls. There are two double walls
\begin{eqnarray}
&&H_{0\la 1 \leftarrow 2 \leftarrow 4\ra}=H_{0\la 1 \leftarrow
2\ra}e^{a_2(r_2)}=\left(
\begin{array}{cccc}
   1 & e^{r_2} & 0 & 0 \\
   0 & 0 & 1 & e^{r_1}\\
  \end{array}
 \right), \no
&&H_{0\la 1 \leftarrow 3 \leftarrow 4\ra}=H_{0\la 1 \leftarrow
3\ra}e^{a_1(r_2)}=\left(
\begin{array}{cccc}
   1 & e^{r_1} & 0 & 0 \\
   0 & 0 & 1 & e^{r_2}\\
  \end{array}
 \right).
  \label{eq:sp2doublewalls}
\end{eqnarray}
The commutation relation (\ref{eq:sp2_commutator}) shows that the
double walls in (\ref{eq:sp2doublewalls}) form a pair of penetrable
walls \cite{Isozumi:2004va} as shown below
\begin{eqnarray}
H_{0\la 1 \leftarrow 2 \leftarrow 4\ra}&=&H_{0\la 1 \leftarrow
2\ra}(r_1)e^{a_2(r_2)}=H_{0\la 1 \ra}e^{a_1(r_1)}e^{a_2(r_2)}\nonumber\\
&=&H_{0\la 1 \ra}e^{a_2(r_2)}e^{a_1(r_1)}=H_{0\la 1 \leftarrow 3\ra}(r_2)e^{a_1(r_1)}\nonumber\\
&=&H_{0\la 1 \leftarrow 3 \leftarrow 4\ra}.
\end{eqnarray}
The walls go through each other as the parameters $r_1$ and $r_2$
vary. This is illustrated in Figure \ref{fig:sp2_double_penetrable}.
%
\subsection{$N=3$ case} \label{sec:sp3}
There are six vacua in $Sp(3)/U(3)$ as shown in \ref{sec:app1}. The
moduli matrices for the vacua (\ref{eq:PhiO6}) can be derived from
the relation (\ref{eq:phi_h}):
\begin{eqnarray}
 &&H_{0\la 1 \ra}=\left(
  \begin{array}{cccccc}
   1 & 0 & 0 & 0 & 0 & 0 \\
   0 & 0 & 1 & 0 & 0 & 0 \\
   0 & 0 & 0 & 0 & 1 & 0 \\
  \end{array}
 \right),~
H_{0\la 2 \ra}=\left(
  \begin{array}{cccccc}
   1 & 0 & 0 & 0 & 0 & 0 \\
   0 & 0 & 1 & 0 & 0 & 0 \\
   0 & 0 & 0 & 0 & 0 & 1 \\
  \end{array}
 \right),\nonumber\\
&&H_{0\la 3 \ra}=\left(
  \begin{array}{cccccc}
   1 & 0 & 0 & 0 & 0 & 0 \\
   0 & 0 & 0 & 1 & 0 & 0 \\
   0 & 0 & 0 & 0 & 1 & 0 \\
  \end{array}
 \right),~
H_{0\la 4 \ra}=\left(
  \begin{array}{cccccc}
   1 & 0 & 0 & 0 & 0 & 0 \\
   0 & 0 & 0 & 1 & 0 & 0 \\
   0 & 0 & 0 & 0 & 0 & 1 \\
  \end{array}
 \right),\nonumber\\
&&H_{0\la 5 \ra}=\left(
  \begin{array}{cccccc}
   0 & 1 & 0 & 0 & 0 & 0 \\
   0 & 0 & 1 & 0 & 0 & 0 \\
   0 & 0 & 0 & 0 & 1 & 0 \\
  \end{array}
 \right),~
H_{0\la 6 \ra}=\left(
  \begin{array}{cccccc}
   0 & 1 & 0 & 0 & 0 & 0 \\
   0 & 0 & 1 & 0 & 0 & 0 \\
   0 & 0 & 0 & 0 & 0 & 1 \\
  \end{array}
 \right),\nonumber\\
&&H_{0\la 7 \ra}=\left(
  \begin{array}{cccccc}
   0 & 1 & 0 & 0 & 0 & 0 \\
   0 & 0 & 0 & 1 & 0 & 0 \\
   0 & 0 & 0 & 0 & 1 & 0 \\
  \end{array}
 \right),~
H_{0\la 8 \ra}=\left(
  \begin{array}{cccccc}
   0 & 1 & 0 & 0 & 0 & 0 \\
   0 & 0 & 0 & 1 & 0 & 0 \\
   0 & 0 & 0 & 0 & 0 & 1 \\
  \end{array}
 \right).
 \label{eq:sp3_vac}
\end{eqnarray}
The plus-minus signs of $\alpha_i$ and $\beta_i$ in (\ref{eq:PhiO6})
have been removed by the world-volume symmetry (\ref{WVS}).

There are three operators generating elementary walls
\begin{eqnarray}
a_1&=&\left(
  \begin{array}{cccccc}
   0 & 0 & 0 & 0 & 0 &0 \\
   0 & 0 & 0 & 0 & 0 &0 \\
   0 & 0 & 0 & 0 & 0 &0 \\
   0 & 0 & 0 & 0 & 0 &0 \\
   0 & 0 & 0 & 0 & 0 &1 \\
   0 & 0 & 0 & 0 & 0 &0 \\
  \end{array}
 \right),~~
a_2=\left(
  \begin{array}{cccccc}
   0 & 0 & 0 & 0 & 0 &0 \\
   0 & 0 & 0 & 0 & 0 &0 \\
   0 & 0 & 0 & 1 & 0 &0 \\
   0 & 0 & 0 & 0 & 0 &0 \\
   0 & 0 & 0 & 0 & 0 &0 \\
   0 & 0 & 0 & 0 & 0 &0 \\
  \end{array}
 \right),
\nonumber
\end{eqnarray}
\begin{eqnarray}
a_3&=&\left(
  \begin{array}{cccccc}
   0 & 1 & 0 & 0 & 0 &0 \\
   0 & 0 & 0 & 0 & 0 &0 \\
   0 & 0 & 0 & 0 & 0 &0 \\
   0 & 0 & 0 & 0 & 0 &0 \\
   0 & 0 & 0 & 0 & 0 &0 \\
   0 & 0 & 0 & 0 & 0 &0 \\
  \end{array}
 \right).
\end{eqnarray}
Each operator generates four elementary walls
\begin{eqnarray}
\begin{array}{ccc}
H_{0\la 1 \leftarrow 2\ra}=H_{0\la 1 \ra}e^{a_1(r_1)},&&
H_{0\la 3 \leftarrow 4\ra}=H_{0\la 3 \ra}e^{a_1(r_1)},\\
H_{0\la 5 \leftarrow 6\ra}=H_{0\la 5 \ra}e^{a_1(r_1)},&&
H_{0\la 7 \leftarrow 8\ra}=H_{0\la 7 \ra}e^{a_1(r_1)},\\
\end{array}\vspace{1cm}\\
\begin{array}{ccc}
H_{0\la 1 \leftarrow 3\ra}=H_{0\la 1\ra}e^{a_2(r_1)},&&
H_{0\la 2 \leftarrow 4\ra}=H_{0\la 2 \ra}e^{a_2(r_1)},\\
H_{0\la 5 \leftarrow 7\ra}=H_{0\la 5 \ra}e^{a_2(r_1)},&&
H_{0\la 6 \leftarrow 8\ra}=H_{0\la 6 \ra}e^{a_2(r_1)},\\
\end{array}\\
\begin{array}{ccc}
H_{0\la 1 \leftarrow 5\ra}=H_{0\la 1 \ra}e^{a_3(r_1)},&&
H_{0\la 2 \leftarrow 6\ra}=H_{0\la 2 \ra}e^{a_3(r_1)},\\
H_{0\la 3 \leftarrow 7\ra}=H_{0\la 3 \ra}e^{a_3(r_1)},&&
H_{0\la 4 \leftarrow 8\ra}=H_{0\la 4 \ra}e^{a_3(r_1)},\\
\end{array}
\end{eqnarray}
where $a_i(r)\equiv e^ra_i$.

We find that all the operators $a_1$, $a_2$ and $a_3$ are
commutative:
\begin{eqnarray}
[a_1, a_2]=[a_1, a_3]=[a_2, a_3]=0. \label{eq:sp6_commutator}
\end{eqnarray}
It shows that there is no compressed wall. All the multiwalls are
thereby penetrable in some intermediate regions. Each vanishing
commutator in (\ref{eq:sp6_commutator}) describes two pairs of
penetrable double walls as follows
\begin{eqnarray}
&&[a_1, a_2]=0 ~~ \Rightarrow ~~ H_{0\la 1 \leftarrow 2,3 \leftarrow
4\ra}, ~~ H_{0\la 5 \leftarrow 6,7 \leftarrow 8\ra}, \nonumber\\
&&[a_1, a_3]=0 ~~ \Rightarrow ~~ H_{0\la 1 \leftarrow 2,5 \leftarrow
6\ra}, ~~ H_{0\la 3 \leftarrow 4,7 \leftarrow 8\ra}, \nonumber\\
&&[a_2, a_3]=0 ~~ \Rightarrow ~~ H_{0\la 1 \leftarrow 3,5 \leftarrow
7\ra}, ~~ H_{0\la 2 \leftarrow 4,6 \leftarrow 8\ra}.
\end{eqnarray}
Here, for instance, $\la 1 \leftarrow 2,3 \leftarrow 4\ra$ means a
configuration passing intermediate vacuum $2$ or $3$ according to a
choice of parameters. The triple wall $H_{0\la 1 \leftarrow 2
\leftarrow 4 \leftarrow 8\ra}$ can be constructed by multiplying the
other elementary-wall operator $e^{a_3(r)}$ to the double wall
$H_{0\la 1 \leftarrow 2,3 \leftarrow 4\ra}=H_{0\la
1\ra}e^{a_1(r)}e^{a_2(r)}$ as
\begin{figure}[h!]
\begin{center}
\epsfxsize=7cm
   \epsfbox{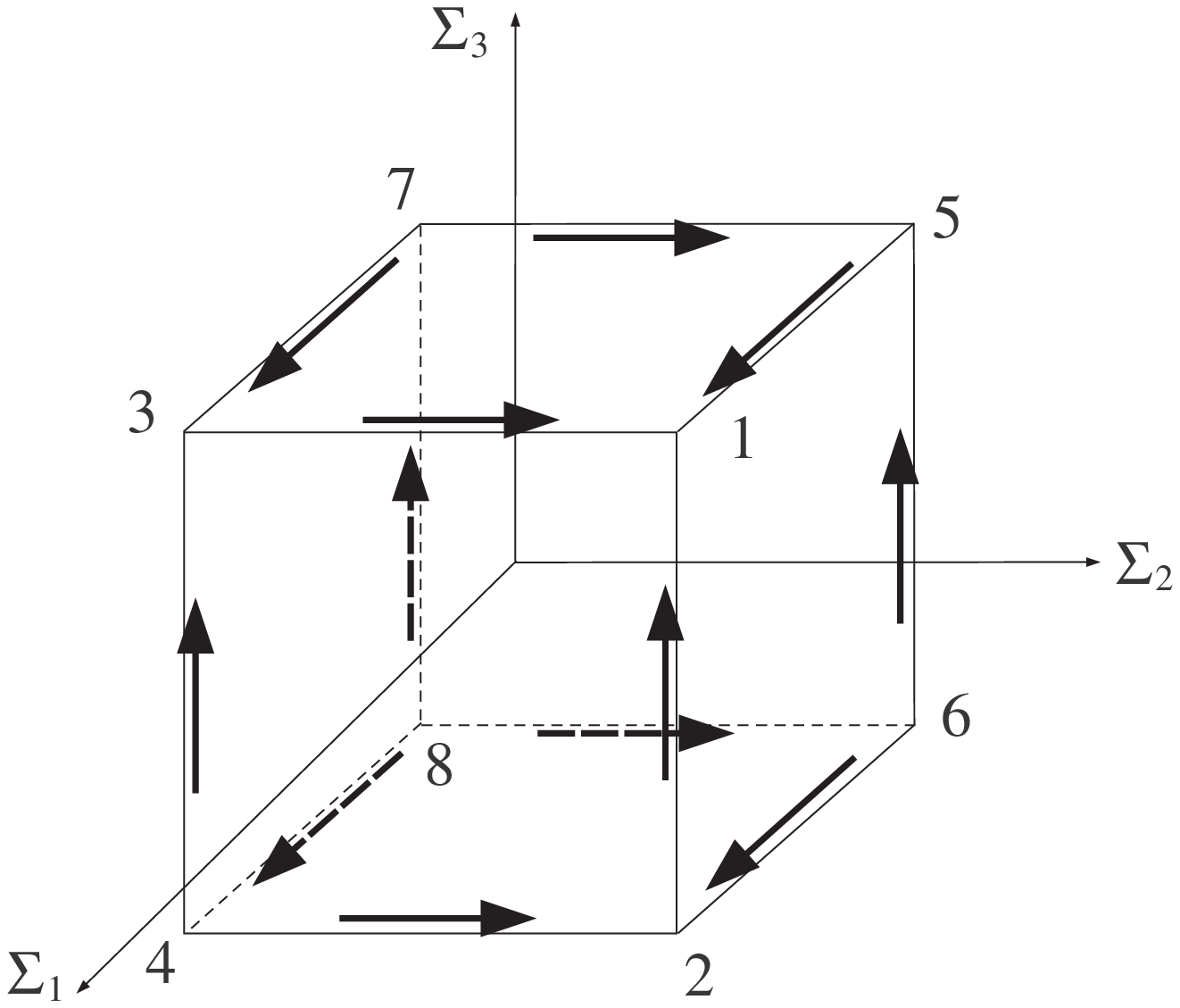}\\
   (a)\\
$\begin{array}{cc}
  \epsfxsize=7cm
   \epsfbox{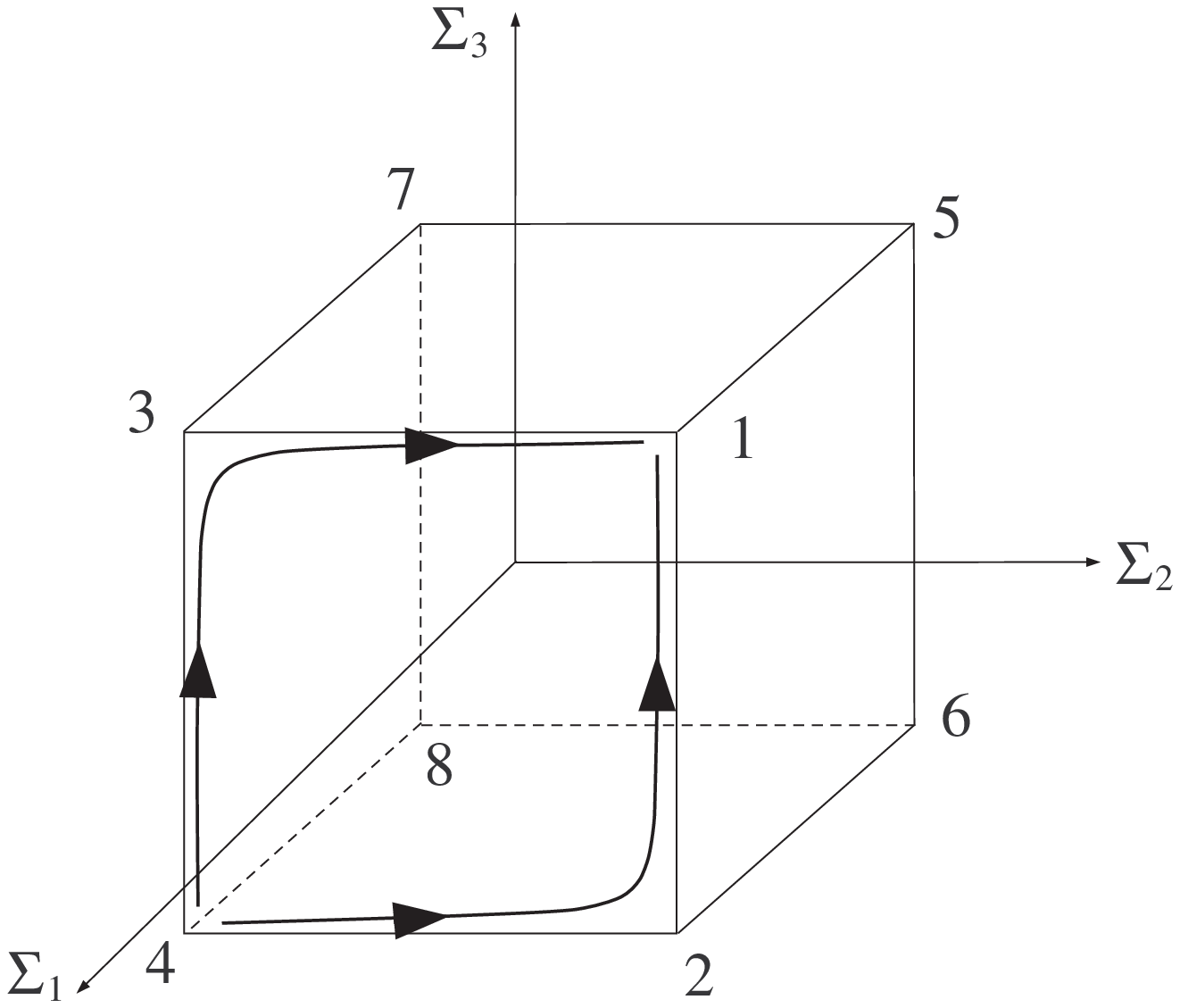}
  &
  \epsfxsize=7cm
   \epsfbox{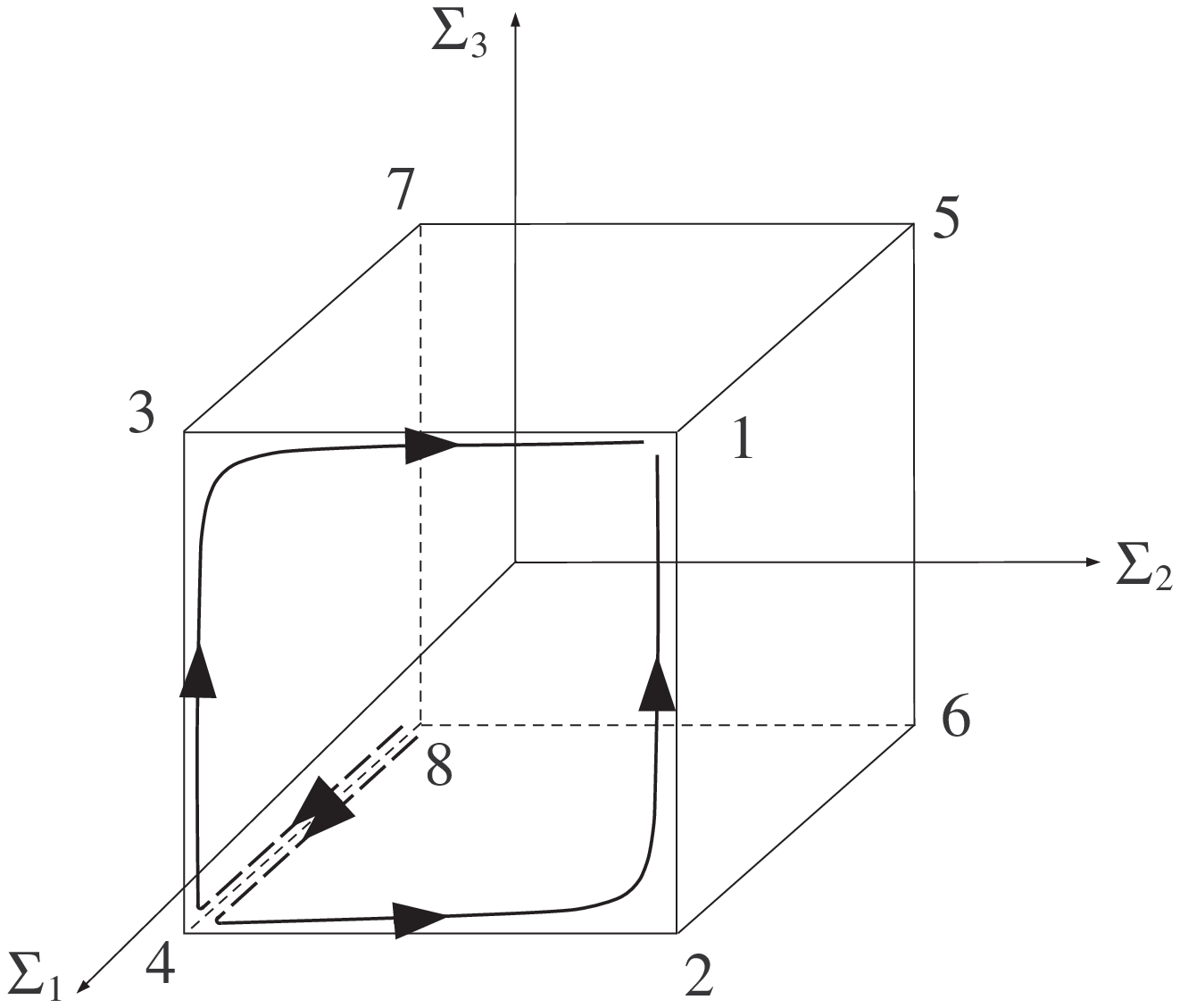}\\
   \mathrm{(b)}&\mathrm{(c)}
\end{array}$
  \caption{ Walls in $Sp(3)/U(3)$. (a)Single walls, which are all elementary walls.
   (b)A pair of double walls $H_{0\la 1 \leftarrow 2,3 \leftarrow 4\ra}$, which are penetrable.
   (c) Triple walls $H_{0\la 1 \leftarrow 2,3 \leftarrow 4\leftarrow 8\ra}$.}
  \label{fig:sp3_single_double_triple}
\end{center}
\end{figure}
\begin{eqnarray}
H_{0\la 1 \leftarrow 2 \leftarrow 4 \leftarrow 8\ra}=H_{0\la
1\ra}e^{a_1(r)}e^{a_2(r)}e^{a_3(r)}. \label{eq:sp6_triple}
\end{eqnarray}
Since all the elementary-wall operators are commutative, the triple
wall (\ref{eq:sp6_triple}) describes all the six triple walls
$H_{0\la 1 \leftarrow 2 \leftarrow 4 \leftarrow 8\ra}$, $H_{0\la 1
\leftarrow 2 \leftarrow 6 \leftarrow 8\ra}$, $H_{0\la 1 \leftarrow 3
\leftarrow 4 \leftarrow 8\ra}$, $H_{0\la 1 \leftarrow 3 \leftarrow 7
\leftarrow 8\ra}$, $H_{0\la 1 \leftarrow 5 \leftarrow 6 \leftarrow
8\ra}$ and $H_{0\la 1 \leftarrow 5 \leftarrow 7 \leftarrow 8\ra}$.
They are penetrable each other in the sense that one or two
intermediate vacua interchange. The triple walls are the multiwalls
composed of the maximal number of single walls in $Sp(3)/U(3)$. The
elementary walls, a pair of double walls $H_{0\la 1 \leftarrow 2,3
\leftarrow 4\ra}$ and triple walls $H_{0\la 1 \leftarrow 2,3
\leftarrow 4\leftarrow 8\ra}$ are drawn in Figure
\ref{fig:sp3_single_double_triple}.

%
%
%
\section{Conclusion}
We have studied the vacuum structure and domain walls interpolating
the vacua in the massive K\"{a}hler NLSM on $SO(2N)/U(N)$ and
$Sp(N)/U(N)$ in three-dimensional spacetime in the moduli matrix
approach. The mass term has been introduced by dimensional reduction
from the massless $\mathcal{N}=1$ K\"{a}hler NLSM in
four-dimensional spacetime.

For $SO(2N)/U(N)$ case, we have found that there exist $2^{N-1}$
discrete vacua. We have studied wall solutions interpolating those
vacua explicitly for $N=2$ and $3$ cases. Elementary walls in
\cite{Isozumi:2004va} are defined by positive step operators
changing the flavor for the same color by one unit in the Grassmann
manifold. In $SO(2N)/U(N)$, however the elementary walls of the
definition are not compatible with the $F$-term constraint, which
defines $SO(2N)$ group. We have modified the formalism in the
Grassmann manifold slightly. We have introduced an additional unit
element in the step operators for elementary walls taking account of
the world-volume symmetry of the moduli matrices for vacua which
correspond to the boundaries of the elementary walls. For $N=2$
case, we have shown that there is only one elementary wall, as
expected from $\mathbf{C}P^1$ model, which is isomorphic to
$SO(4)/U(2)$. For $N=3$ case, we have obtained the wall solutions up
to triple walls. We have compared the result with $\mathbf{C}P^3$,
which is isomorphic to $SO(6)/U(3)$. From this, we have shown that
the method used for $SO(2N)/U(N)$ does not compromise the algebras
of operators generating walls.

For $Sp(N)/U(N)$ case, we have found that there exist $2^{N}$
discrete vacua. We have studied wall solutions interpolating those
vacua for $N=1,2$ and $3$ cases. The formalism in
\cite{Isozumi:2004va} has been consistently applied to the models in
the presence of the $F$-term constraint in this case. We have shown
that there is a single elementary wall for $N=1$ case. We have also
shown that there exist multiwalls up to double walls and triple
walls for $N=2$ and $N=3$ cases, respectively.
%
%
\vspace{8mm}
\section*{Acknowledgments}
We would like to thank Sunggeun Lee for a collaboration at an early
stage. We thank Minoru Eto, Hiroaki Nakajima, Muneto Nitta and
Norisuke Sakai for useful discussion. The work of M.~A. was supported
in part by the Research Program MSM6840770029 and by the project of
International Cooperation ATLAS-CERN of the Ministry of Education,
Youth and Sports of the Czech Republic.

%
%
\appendix
\def\thesection{Appendix \Alph{section}}
\setcounter{equation}{0}
\renewcommand{\theequation}{\Alph{section}.\arabic{equation}}
\section{Vacuum structure of $N=3$ case}\label{sec:app1}
In this Appendix, we derive the vacuum structure for $N=3$ case.
There are eight vacua according to (\ref{vac_sigma2}):
\begin{eqnarray}
&& \Phi_{\la 1 \ra}=\left(
  \begin{array}{cccccc}
   1 & 0 & 0 & 0 & 0 & 0 \\
   0 & 0 & \alpha_1 & 0 & 0 & 0 \\
   0 & 0 & 0 & 0 & \beta_1 & 0 \\
  \end{array}
 \right),~~~(\Sigma_1,\Sigma_2,\Sigma_3)=(m_1,m_2,m_3),\nonumber\\
&&\Phi_{\la 2 \ra}=\left(
  \begin{array}{cccccc}
   1 & 0 & 0 & 0 & 0 & 0 \\
   0 & 0 & \alpha_2 & 0 & 0 & 0 \\
   0 & 0 & 0 & 0 & 0 & \beta_2 \\
  \end{array}
 \right),~~~(\Sigma_1,\Sigma_2,\Sigma_3)=(m_1,m_2,-m_3),\nonumber\\
&&\Phi_{\la 3 \ra}=\left(
  \begin{array}{cccccc}
   1 & 0 & 0 & 0 & 0 & 0 \\
   0 & 0 & 0 & \alpha_3 & 0 & 0 \\
   0 & 0 & 0 & 0 & \beta_3 & 0 \\
  \end{array}
 \right),~~~(\Sigma_1,\Sigma_2,\Sigma_3)=(m_1,-m_2,m_3),\nonumber\\
&&\Phi_{\la 4 \ra}=\left(
  \begin{array}{cccccc}
   1 & 0 & 0 & 0 & 0 & 0 \\
   0 & 0 & 0 & \alpha_4 & 0 & 0 \\
   0 & 0 & 0 & 0 & 0 & \beta_4 \\
  \end{array}
 \right),~~~(\Sigma_1,\Sigma_2,\Sigma_3)=(m_1,-m_2,-m_3),\nonumber\\
&&\Phi_{\la 5 \ra}=\left(
  \begin{array}{cccccc}
   0 & 1 & 0 & 0 & 0 & 0 \\
   0 & 0 & \alpha_5 & 0 & 0 & 0 \\
   0 & 0 & 0 & 0 & \beta_5 & 0 \\
  \end{array}
 \right),~~~(\Sigma_1,\Sigma_2,\Sigma_3)=(-m_1,m_2,m_3),\nonumber\\
&&\Phi_{\la 6 \ra}=\left(
  \begin{array}{cccccc}
   0 & 1 & 0 & 0 & 0 & 0 \\
   0 & 0 & \alpha_6 & 0 & 0 & 0 \\
   0 & 0 & 0 & 0 & 0 & \beta_6 \\
  \end{array}
 \right),~~~(\Sigma_1,\Sigma_2,\Sigma_3)=(-m_1,m_2,-m_3),\nonumber\\
&&\Phi_{\la 7 \ra}=\left(
  \begin{array}{cccccc}
   0 & 1 & 0 & 0 & 0 & 0 \\
   0 & 0 & 0 & \alpha_7 & 0 & 0 \\
   0 & 0 & 0 & 0 & \beta_7 & 0 \\
  \end{array}
 \right),~~~(\Sigma_1,\Sigma_2,\Sigma_3)=(-m_1,-m_2,m_3),\nonumber\\
&&\Phi_{\la 8 \ra}=\left(
  \begin{array}{cccccc}
   0 & 1 & 0 & 0 & 0 & 0 \\
   0 & 0 & 0 & \alpha_8 & 0 & 0 \\
   0 & 0 & 0 & 0 & 0 & \beta_8 \\
  \end{array}
 \right),~~~(\Sigma_1,\Sigma_2,\Sigma_3)=(-m_1,-m_2,-m_3).
 \label{eq:PhiO6}
\end{eqnarray}
where $\alpha_i=1$ and $\beta_i=1 (i=1,\cdots,8)$ for $SO(6)/U(3)$
whereas $\alpha_i=\pm1$ and $\beta_i=\pm1$ for $Sp(3)/U(3)$.

For $\epsilon=+1$, the half of them are parity-related to the other
half as in $N=2$ case. We define rotation transformations
${\mathcal{R}}_i~(i=1,2,3)$
\begin{eqnarray}
{\mathcal{R}}_1 =\left(
\begin{array}{ccc}
I & 0 & 0 \\
0 & P & 0 \\
0 & 0 & P \\
\end{array}
\right),~~
{\mathcal{R}}_2 =\left(
\begin{array}{ccc}
P & 0 & 0 \\
0 & I & 0 \\
0 & 0 & P \\
\end{array}
\right),~~
{\mathcal{R}}_3 =\left(
\begin{array}{ccc}
P & 0 & 0 \\
0 & P & 0 \\
0 & 0 & I \\
\end{array}
\right),
\end{eqnarray}
and a parity transformation of $O(6)$ group,
\begin{eqnarray}
{\mathcal{P}}=\left(
\begin{array}{ccc}
I & 0 & 0 \\
0 & I & 0 \\
0 & 0 & P \\
\end{array}
\right),
\end{eqnarray}
where $I$ and $P$ are the same as defined by (\ref{eq:PhiO4}). The
eight vacua (\ref{eq:PhiO6}) are related by the rotational
transformations as
\begin{eqnarray}
\Phi_{\la 1 \ra}=\Phi_{\la 4 \ra}{\mathcal{R}}_1,~~
\Phi_{\la 6 \ra}=\Phi_{\la 7 \ra}{\mathcal{R}}_1, \nonumber\\
\Phi_{\la 1 \ra}=\Phi_{\la 6 \ra}{\mathcal{R}}_2,~~
\Phi_{\la 4 \ra}=\Phi_{\la 7 \ra}{\mathcal{R}}_2, \nonumber\\
\Phi_{\la 1 \ra}=\Phi_{\la 7 \ra}{\mathcal{R}}_3,~~ \Phi_{\la 4
\ra}=\Phi_{\la 6 \ra}{\mathcal{R}}_3,
\end{eqnarray}
while the vacua $\Phi_{\la 1 \ra}$ and $\Phi_{\la 2 \ra}$ are related by the parity
transformation as
\begin{eqnarray}
\Phi_{\la 1 \ra}=\Phi_{\la 2 \ra}{\mathcal{P}}.
\end{eqnarray}
It shows that the vacua $\Phi_{\la 1 \ra}$, $\Phi_{\la 4 \ra}$,
$\Phi_{\la 6 \ra}$ and $\Phi_{\la 7 \ra}$ are in one $SO(6)/U(3)$
manifold and the rest of the vacua are in the other $SO(6)/U(3)$
manifold. Focusing on one of the manifolds, we obtain four discrete
vacua.

For $\epsilon=-1$, all the vacua (\ref{eq:PhiO6}) are in a single
$Sp(3)/U(3)$ model since the constraint (\ref{eq:constraint2}) is an
invariant submanifold under the action of $Sp(3)$ group. We
therefore find that there exist $2^3=8$ discrete vacua in the
$Sp(3)/U(3)$ model.
%
\section{${\mathbf{C}}P^3$}\label{sec:app2}
\setcounter{equation}{0}
In this Appendix, we study the vacuum structure and domain walls
interpolating them in a massive K\"ahler NLSM on ${\mathbf{C}}P^3$.
The action is given by (\ref{3d-lag}) with $U(1)$ gauge symmetry and
vanishing $F$-term constraint
\begin{eqnarray}
{\mathcal{L}}_{\mathrm{bos}~3D}&=&-|D_m\phi^i|^2-|i\phi^j
M_j^{~i}-i\Sigma \phi^i|^2
+|F^i|^2+\frac{1}{2}D(\phi^i\bar{\phi}_i-1).
\end{eqnarray}
The mass matrix $M_i^{~j}$ is given by a linear combination of
Cartan matrices for $SU(4)$
\begin{eqnarray}
{1 \over \sqrt{2}}{\rm diag}(1, -1, 0, 0),~~~{1 \over \sqrt{6}}{\rm
diag}(1, 1, -2, 0),~~~{1 \over 2 \sqrt{3}}{\rm diag}(1, 1, 1, -3),
\end{eqnarray}
leading to
\begin{eqnarray}
 M={\rm diag}(m_1', m_2', m_3', m_4'),
\end{eqnarray}
where
\begin{eqnarray}
m^\prime_1&=&m_1+m_2+m_3, \nonumber\\
m^\prime_2&=&-m_1+m_2+m_3, \nonumber\\
m^\prime_3&=&-2m_2+m_3, \nonumber\\
m^\prime_4&=&-3m_3.
\end{eqnarray}
The mass parameters have been scaled as ${1 \over
\sqrt{2}}m_1\rightarrow m_1$, ${-1 \over \sqrt{2}}m_2\rightarrow
m_2$, ${-2 \over \sqrt{6}}m_3\rightarrow m_3$ and ${1 \over
2\sqrt{3}}m_4\rightarrow m_4$.

Let us see the vacuum structure and domain wall solutions. There are
four vacua of which the moduli matrices are
\begin{eqnarray}
&&H_{0\la 1 \ra}=(1,0,0,0),~~~\Sigma=m^\prime_1,\nonumber\\
&&H_{0\la 2 \ra}=(0,1,0,0),~~~\Sigma=m^\prime_2,\nonumber\\
&&H_{0\la 3 \ra}=(0,0,1,0),~~~\Sigma=m^\prime_3,\nonumber\\
&&H_{0\la 4 \ra}=(0,0,0,1),~~~\Sigma=m^\prime_4.
\end{eqnarray}
We assume that $m^\prime_i>m^\prime_{i+1}$. Operators generating
elementary walls are obtained as
\begin{eqnarray}
\begin{array}{ccccc}
a^\prime_{1}=\left(
  \begin{array}{cccc}
   0 & 1 & 0 & 0 \\
   0 & 0 & 0 & 0 \\
   0 & 0 & 0 & 0 \\
   0 & 0 & 0 & 0 \\
   \end{array}
 \right),~~
a^\prime_{2}=\left(
  \begin{array}{cccc}
   0 & 0 & 0  & 0 \\
   0 & 0 & 1  & 0 \\
   0 & 0 & 0  & 0 \\
   0 & 0 & 0  & 0 \\
   \end{array}
 \right),~~
a^\prime_{3}=\left(
  \begin{array}{cccccc}
   0 & 0 & 0 & 0 \\
   0 & 0 & 0 & 0 \\
   0 & 0 & 0 & 1 \\
   0 & 0 & 0 & 0 \\
   \end{array}
 \right).
 \label{eq:cp3_creation_op}
\end{array}
\end{eqnarray}
The operators (\ref{eq:cp3_creation_op}) generate three elementary
walls
\begin{eqnarray}
H_{0\la 1 \leftarrow 2\ra}&=&H_{0\la 1 \ra}e^{a^\prime_1(r)}=(1,e^r,0,0),\nonumber\\
H_{0\la 2 \leftarrow 3\ra}&=&H_{0\la 2 \ra}e^{a^\prime_2(r)}=(0,1,e^r,0),\nonumber\\
H_{0\la 3 \leftarrow 4\ra}&=&H_{0\la 3
\ra}e^{a^\prime_3(r)}=(0,0,1,e^r).
\end{eqnarray}
Non-vanishing commutators of the matrices in
(\ref{eq:cp3_creation_op}) are
\begin{eqnarray}
E^\prime_1=[a^\prime_1,a^\prime_2]\neq0,~~~E^\prime_2=[a^\prime_2,a^\prime_3]\neq0,
\label{eq:cp3_gen_comp1}
\end{eqnarray}
which generate compressed walls of level one
\begin{eqnarray}
&&H_{0\la 1 \leftarrow 3\ra}=H_{0\la 1 \ra}e^{E^\prime_1}=(1,0,e^r,0),\\
&&H_{0\la 2 \leftarrow 4\ra}=H_{0\la 2
\ra}e^{E^\prime_2}=(0,1,0,e^r).
\end{eqnarray}
The commutators among operators generating elementary walls and
compressed walls of level one are
\begin{eqnarray}
E^\prime_3=[a^\prime_1,E^\prime_2]=[E^\prime_1,a^\prime_3]\neq0,
\label{eq:cp3_gen_comp2}
\end{eqnarray}
which generates a compressed wall of level two
\begin{eqnarray}
H_{0\la 1 \leftarrow 4\ra}=H_{0\la 1 \ra}e^{E^\prime_3}=(1,0,0,e^r).
\end{eqnarray}
We give moduli matrices of multiwalls. Double walls are
\begin{eqnarray}
&&H_{0\la 1 \leftarrow 2 \leftarrow 3\ra}=H_{0\la 1 \leftarrow 2\ra}e^{a^\prime_2(r_2)}=(1,e^{r_1},e^{r_1+r_2},0),\nonumber\\
&&H_{0\la 2 \leftarrow 3 \leftarrow 4\ra}=H_{0\la 2 \leftarrow 3\ra}e^{a^\prime_3(r_2)}=(0,1,e^{r_1},e^{r_1+r_2}),\nonumber\\
&&H_{0\la 1 \leftarrow 2 \leftarrow 4\ra}=H_{0\la 1 \leftarrow 2\ra}e^{E^\prime_2(r_2)}=(1,e^{r_1},0,e^{r_1+r_2}),\nonumber\\
&&H_{0\la 1 \leftarrow 3 \leftarrow 4\ra}=H_{0\la 1 \leftarrow
3\ra}e^{a^\prime_3(r_2)}=(1,0,e^{r_1},e^{r_1+r_2}).
\end{eqnarray}
There is one triple wall
\begin{eqnarray}
H_{0\la 1 \leftarrow 2 \leftarrow 3 \leftarrow 4\ra}=H_{0\la 1
\leftarrow 2 \leftarrow
3\ra}e^{a^\prime_3(r_3)}=(1,e^{r_1},e^{r_1+r_2},e^{r_1+r_2+r_3}).
\end{eqnarray}

We compare the vacuum structure and the algebras of operators of
$SO(6)/U(3)$ in Section \ref{sec:so6} with the result of
$\mathbf{C}P^3$. The number of vacua and the number of generating
operators for elementary walls of $SO(6)/U(3)$ are the same as each
of ${\mathbf{C}}P^3$. The algebras for compressed walls
(\ref{eq:so6_gen_comp1}) and (\ref{eq:so6_gen_comp2}) are also the
same as (\ref{eq:cp3_gen_comp1}) and (\ref{eq:cp3_gen_comp2}). The
vacuum structure and wall configuration of $SO(6)/U(3)$ coincide
with those of ${\mathbf{C}}P^3$ under the relabeling of the vacua
$\la 1 \ra$ to $\la 1 \ra$, $\la 4 \ra$ to $\la 2 \ra$, $\la 6 \ra$
to $\la 3 \ra$ and $\la 7 \ra$ to $\la 4 \ra$.
%
%
%


\begin{thebibliography}{99}
%
\bibitem{Witten:1978mh}
  E.~Witten and D.~I.~Olive,
  Phys.\ Lett.\  B {\bf 78}, 97 (1978).
%
\bibitem{BPS}
 E.~Bogomol'nyi,
  {\it Sov.\ J.\ Nucl.\ Phys.\ } {\bf B24} (1976) 449;
 M.~K.~Prasad and C.~H.~Sommerfield,
  {\it Phys.\ Rev.\ Lett.\ } {\bf 35}, 760 (1975).
%
\bibitem{de Azcarraga:1989gm}
  J.~A.~de Azcarraga, J.~P.~Gauntlett, J.~M.~Izquierdo and P.~K.~Townsend,
  Phys.\ Rev.\ Lett.\  {\bf 63}, 2443 (1989).

\bibitem{CQR}
  M.~Cvetic, F.~Quevedo and S.~J.~Rey,
   Phys.\ Rev.\ Lett.\  {\bf 67}, 1836 (1991);
  M.~Cvetic, S.~Griffies and S.~J.~Rey,
   Nucl.\ Phys.\  B {\bf 381}, 301 (1992)
   [arXiv:hep-th/9201007].
%
\bibitem{Abraham:1992vb}
  E.~R.~C.~Abraham and P.~K.~Townsend,
  Phys.\ Lett.\  B {\bf 291}, 85 (1992).
%
\bibitem{Isozumi:2004jc}
  Y.~Isozumi, M.~Nitta, K.~Ohashi and N.~Sakai,
  Phys.\ Rev.\ Lett.\  {\bf 93}, 161601 (2004)
  [arXiv:hep-th/0404198].
%
\bibitem{Isozumi:2004va}
  Y.~Isozumi, M.~Nitta, K.~Ohashi and N.~Sakai,
  Phys.\ Rev.\  D {\bf 70}, 125014 (2004)
  [arXiv:hep-th/0405194].
%
\bibitem{INOS3}
  Y.~Isozumi, M.~Nitta, K.~Ohashi and N.~Sakai,
  Phys.\ Rev.\  D {\bf 71}, 065018 (2005)
  [arXiv:hep-th/0405129].
%
\bibitem{EINOS1}
  M.~Eto, Y.~Isozumi, M.~Nitta, K.~Ohashi and N.~Sakai,
  Phys.\ Rev.\  D {\bf 72}, 085004 (2005)
  [arXiv:hep-th/0506135];
  Phys.\ Lett.\  B {\bf 632}, 384 (2006)
  [arXiv:hep-th/0508241].
%
\bibitem{EINOS2}
  M.~Eto, Y.~Isozumi, M.~Nitta, K.~Ohashi and N.~Sakai,
  Phys.\ Rev.\ Lett.\  {\bf 96}, 161601 (2006)
  [arXiv:hep-th/0511088];
  M.~Eto, T.~Fujimori, Y.~Isozumi, M.~Nitta, K.~Ohashi, K.~Ohta and N.~Sakai,
  Phys.\ Rev.\  D {\bf 73}, 085008 (2006)
  [arXiv:hep-th/0601181].
%
\bibitem{EINOS3}
M.~Eto, Y.~Isozumi, M.~Nitta, K.~Ohashi and N.~Sakai,
  Phys.\ Rev.\  D {\bf 72}, 025011 (2005)
  [arXiv:hep-th/0412048].
%
\bibitem{ENOT}
M.~Eto, M.~Nitta, K.~Ohashi and D.~Tong,
  Phys.\ Rev.\ Lett.\  {\bf 95}, 252003 (2005)
  [arXiv:hep-th/0508130].
%
\bibitem{INOS2}
  M.~Eto, Y.~Isozumi, M.~Nitta, K.~Ohashi and N.~Sakai,
  J.\ Phys.\ A  {\bf 39}, R315 (2006)
  [arXiv:hep-th/0602170].
%
\bibitem{ANS}
  M.~Arai, M.~Nitta and N.~Sakai,
  Prog.\ Theor.\ Phys.\  {\bf 113}, 657 (2005)
  [arXiv:hep-th/0307274].
%
\bibitem{Gauntlett:2000bd}
  J.~P.~Gauntlett, D.~Tong and P.~K.~Townsend,
  Phys.\ Rev.\  D {\bf 63}, 085001 (2001)
  [arXiv:hep-th/0007124].
%
\bibitem{Gauntlett:2000de}
  J.~P.~Gauntlett, R.~Portugues, D.~Tong and P.~K.~Townsend,
  Phys.\ Rev.\  D {\bf 63}, 085002 (2001)
  [arXiv:hep-th/0008221].
%
\bibitem{Arai:2002xa}
  M.~Arai, M.~Naganuma, M.~Nitta and N.~Sakai,
  Nucl.\ Phys.\  B {\bf 652}, 35 (2003)
  [arXiv:hep-th/0211103].
%
\bibitem{Arai:2003my}
  M.~Arai, E.~Ivanov and J.~Niederle,
  Nucl.\ Phys.\  B {\bf 680}, 23 (2004)
  [arXiv:hep-th/0312037].
%
\bibitem{KG1}
  S.~J.~Gates,~Jr. and S.~M.~Kuzenko,
  Nucl.\ Phys.\ B {\bf 543}, 122 (1999)
  [arXiv:hep-th/9810137].
%
\bibitem{KG2}
  S.~J.~Gates,~Jr. and S.~M.~Kuzenko,
  Fortsch.\ Phys.\  {\bf 48}, 115 (2000)
  [arXiv:hep-th/9903013].
%
\bibitem{AKL1}
  M.~Arai, S.~M.~Kuzenko and U.~Lindstr\"om,
  JHEP {\bf 0702}, 100 (2007)
  [arXiv:hep-th/0612174].
%
\bibitem{AN}
  M.~Arai and M.~Nitta,
  Nucl.\ Phys.\ B {\bf 745}, 208 (2006)
  [arXiv:hep-th/0602277].
%
\bibitem{AKL2}
  M.~Arai, S.~M.~Kuzenko and U.~Lindstr\"om,
  JHEP {\bf 0712}, 008 (2007)
  [arXiv:0709.2633 [hep-th]].
%
\bibitem{KLR}
  A.~Karlhede, U.~Lindstr\"om and M.~Ro\v cek,
  Phys.\ Lett.\ B {\bf 147}, 297 (1984).
%
\bibitem{LR}
  U.~Lindstr\"om and M.~Ro\v{c}ek,
  Commun.\ Math.\ Phys.\  {\bf 115}, 21 (1988).
%
\bibitem{HN}
  K.~Higashijima and M.~Nitta,
  Prog.\ Theor.\ Phys.\  {\bf 103}, 635 (2000)
  [arXiv:hep-th/9911139].
%
\bibitem{WB}
  J.~Wess and J.~Bagger,
  {\it Supersymmetry and supergravity},
{\rm  Princeton, USA: Univ. Pr. (1992).}
%
\bibitem{Higashijima:2001vk}
  K.~Higashijima, T.~Kimura and M.~Nitta,
  Nucl.\ Phys.\  B {\bf 623}, 133 (2002)
  [arXiv:hep-th/0108084].
%
\bibitem{Arai:2003tc}
  M.~Arai, M.~Nitta and N.~Sakai,
  Prog.\ Theor.\ Phys.\  {\bf 113}, 657 (2005)
  [arXiv:hep-th/0307274].
\end{thebibliography}
\end{document}